\documentclass[
reprint,
superscriptaddress,
 amsmath,amssymb,
pra,
]{revtex4-2}

\usepackage{amsmath}
\usepackage{braket}
\usepackage{dsfont}
\usepackage{float}%don't move picture
\usepackage{graphicx}% Include figure files
\usepackage[hidelinks]{hyperref}
\usepackage{todonotes}
\renewcommand{\vec}[1]{\boldsymbol{#1}}

\begin{document}

% \preprint{APS/123-QED}

\title{Perturbative solution approach for computing the two-photon Kapitza-Dirac effect in a Gaussian beam standing light wave}
\author{Sven Ahrens}
\thanks{ahrens@shnu.edu.cn}%
\author{Chong Zhang}
\author{Ping Ge}
\author{Guweiyi Li}
\author{Baifei Shen}%
\email{bfshen@shnu.edu.cn}
\affiliation{Shanghai Normal University, Shanghai 200234, China}%

\begin{abstract}
Theoretical spin properties of the Kapitza-Dirac effect beyond the plane-wave description are not known in detail. We develop a method for computing electron diffraction of the two-photon Kapitza-Dirac effect in a two-dimensional Gaussian beam standing light wave within a relativistic formulation. The solutions are computed on the basis of time-dependent perturbation theory, where a momentum space formulation with the use of a Fourier transformation of the external potential allows for the solving the perturbative time-integrals. An iteration over each possible quantum state combination leads to a quadratic scaling of our method with respect to spacial grid resolution, where time-stepping does not occur in the numeric implementation. The position- and momentum space grids are adapted to the two-photon interaction geometry at low resolution, for which our study only finds partial convergence of the simulated diffraction pattern. Further, the method has the advantage of having an easy implementable parallelization layout.
\end{abstract}

\maketitle

\section{\label{sec:level1} Introduction}

In 1933 Kapitza and Dirac proposed the, for the time being, exotic idea to diffract electrons at a standing light waves \cite{kapitza_dirac_1933_proposal}, which nevertheless was established experimentally around the millennium change in different dynamical regimes \cite{Bucksbaum_1988_electron_diffraction_regime,Freimund_Batelaan_2001_KDE_first,Freimund_Batelaan_2002_KDE_detection_PRL}. The experimental demonstration of the possibility to observe the effect raised the question, whether electron spin interactions are possible in Kapitza-Dirac scattering \cite{Batelaan_2003_MSGE}. Such spin dynamics is predicted to be possible \cite{ahrens_bauke_2012_spin-kde,ahrens_bauke_2013_relativistic_KDE,McGregor_Batelaan_2015_two_color_spin,dellweg_mueller_2016_interferometric_spin-polarizer,dellweg_mueller_extended_KDE_calculations,ahrens_2017_spin_filter}, within experimental reach \cite{ahrens_2020_two_photon_bragg_scattering}. While laser based electron spin manipulations are discussed for a variety of setups \cite{Karlovets_2011_radiative_polarization,Del_Seipt_Blackburn_Kirk_2017_electron_spin_polarization,Wen_Tamburini_Keitel_2019_polarized_kilo_ampere_beams,van_Kruining_Mackenroth_2019_magnetic_field_polarization,Chen_Keitel_2019_polarized_positrons,Li_Chen_2020_polarized_positron_electron,Xue_Chen_Hatsagortsyan_Keitel_Li_2022_GeV_polarization,Gong_Hatsagortsyan_Keitel_2023_plasma_instability_polarization}, the spin-dependent Kapitza-Dirac effect promises not only the implementation of spin polarization, but also the observation of the electron spin state on the basis of fundamental light-electron interaction only, similar as in the Stern-Gerlach experiment \cite{Stern_Gerlach_1922_1,Stern_Gerlach_1922_2,Stern_Gerlach_1922_3}.

In the past, theoretical descriptions of the Kapitza-Dirac effect were carried out by assuming a plane-wave laser beam. Only recently, the influence of a longitudinal laser beam polarization component on Kapitza-Dirac scattering was studied by us \cite{ahrens_guan_2022_beam_focus_longitudinal}. The longitudinal component does, due to its transverse anti-symmetric functional behavior in space, only exist for non-plane-wave electro-magnetic fields in the source free Maxwell equations. We therefore have attempted a perturbative solution of the spin-dependent diffraction process in \cite{ahrens_guan_2022_beam_focus_longitudinal}, in which the external field was modeling Gaussian beams in terms of a plane-wave superposition approximation. In the present work, we study an extended implementation of the perturbative calculation, in which the plane-wave expansion of the Gaussian beam is obtained in a systematic manner by a Fourier transform of the field. We point out in this context, that time-dependent perturbation theory can be solved in a straight forward manner, for the case of an interaction of energy-eigensolutions of a quantum particle with plane-wave fields.

In comparison with other computational schemes for computing the time-evolution of quantum states, the proposed perturbative method does not involve any time-stepping, which is beneficial for the envisaged relativistic solutions of the Dirac equation. Also, the computation of the perturbative quantities can be split into smaller, independent computational tasks almost arbitrary, which is advantageous for parallelizing the calculation on larger computer clusters. A drawback of our method is a bad scaling with spacial grid resolution, as as discussed in section \ref{sec:execution_scaling}. We thus assume Bragg scattering of the electron in the so-called Bragg regime \cite{batelaan_2000_KDE_first,batelaan_2007_RMP_KDE} and isolate the interaction peak of the Gaussian beam in momentum space, as explained in sections \ref{sec:gaussian_beam_potentials} and \ref{sec:perturbation_theory}. We therefore execute the code at low spacial resolutions and achieve only partial convergence for our method in space, along the $y$-direction, see appendix \ref{sec:code_convergence} for details.

The article is structured as follows. In section \ref{sec:theory_development} we develop a theoretical formalism for our perturbative method. We start this section by a note about the unit system (section \ref{sec:unit_system}), then introduce the computation grid in position- and momentum space (section \ref{sec:simulation_grid}) with a notion for the Dirac bra-ket representation (section \ref{eq:wave_function_representation}) and introducing the Dirac equation in the formalism (section \ref{sec:relativistic_quantum_dynamics}). We then discuss Gaussian beams on the grid (section \ref{sec:gaussian_beam_potentials}) and formulate the application of time-dependent perturbation theory in our framework (section \ref{sec:perturbation_theory}). Further implementation details are discussed in section \ref{sec:implementation_details} by denoting the formal time-evolution (section \ref{sec:time_propagation}), mentioning properties of the standing light-wave in the Kapitza-Dirac effect in the context of the time-integration in perturbation theory (section \ref{sec:laser_field_time_integration}), describing the iteration scheme of our perturbative algorithm (section \ref{sec:perturbative_iteration_sceme}) and introducing the spin orientation for the initial and final quantum state of the electron (section \ref{sec:spin_orientiation}). We conclude with discussing the computation results in section \ref{sec:results}, were we first describe the simulation and its parameters (section \ref{sec:simulation_parameters}), interpret the physical meaning of the results (section \ref{sec:physical_meaning}), consider the role of the negative solutions of the Dirac equation in the context of our numeric scheme (section \ref{sec:negative_solutions}) and add considerations about the code execution time scaling (section \ref{sec:execution_scaling}). We finally summarize our study and give an outlook about further developing the method in section \ref{sec:discussion_and_outlook}. We supplement the article with an appendix about the solutions for the perturbative time-integral (appendix \ref{sec:perturbative_integral_solution}), discussing the code convergence of our method (section \ref{sec:code_convergence}) and investigating the influence of the longitudinal laser polarization component on the quantum dynamics (section \ref{sec:further_properties}).

\section{Theoretical model: Introduction of numeric perturbation approach\label{sec:theory_development}}

\subsection{Unit system\label{sec:unit_system}}

In our article, we are denoting the reduced Planck constant $\hbar$, the electron rest mass $m$, the vacuum speed of light $c$ and the electron charge $e$, in the context of a Gaussian unit system.

\subsection{Simulation grid and geometry\label{sec:simulation_grid}}

We want to implement our description on a two-dimensional grid with coordinates
\begin{subequations}\label{eq:position_space_grid}%
\begin{align}%
x_{m}&=x_\textrm{min}+m\,\Delta x \,,\quad m \in \{ 0,\dots,N_{x}-1  \}  \\
y_{n}&=y_\textrm{min}+n\,\Delta y \,,\quad n \in \{ 0,\dots,N_{y}-1  \}
\end{align}%
\end{subequations}%
in position space and
\begin{subequations}\label{eq:momentum_space_grid}%
\begin{align}%
k_{x,a}&=\kappa_x+a\,\Delta k_x \,,\quad a \in \{ 0,\dots,N_{x}-1  \}  \label{eq:momentum_space_grid_x}\\
k_{y,b}&=\kappa_y+b\,\Delta k_y \,,\quad b \in \{ 0,\dots,N_{y}-1  \}
\end{align}%
\end{subequations}%
in momentum space. For mathematical and physical reasons, extensions of the grid space arrangement in particular for the electron quantum state need to be formulated later, in section \ref{sec:perturbation_theory}. In Eqs. \eqref{eq:position_space_grid} and \eqref{eq:momentum_space_grid} we have introduced the grid spacing $\Delta x$, $\Delta y$ in position space, $\Delta k_x$, $\Delta k_y$ in momentum space, the offset coordinates $x_\textrm{min}$, $y_\textrm{min}$ in position space and $\kappa_x$, $\kappa_y$ in momentum space, with the grid resolution $N_x$, $N_y$. We mention that for technical reasons, the grid resolution is always odd valued in this article. The coordinates \eqref{eq:position_space_grid} and \eqref{eq:momentum_space_grid} can be written as
\begin{subequations}%
\begin{align}%
\vec x_{m,n}&=\vec x_\textrm{min}+m\,{\Delta}x\,\vec e_x+n\,{\Delta}y\,\vec e_{y}\\
\vec{k}_{a,b}^{\vec{\kappa}}&=\vec{\kappa}+a\,{\Delta}k_{x}\,\vec{e_{x}}+b\,{\Delta}k_{y}\,\vec{e_{y}}
\end{align}%
\end{subequations}%
in shorthand vector notion, with the orthogonal unit vectors $\vec e_x$, $\vec e_y$ along the coordinate axes. Note, that we are adding the index $\vec \kappa$ in the wave vector $\vec{k}_{a,b}^{\vec{\kappa}}$ to have a flexible notion for multiple momentum offsets in one system.

If we impose the condition
\begin{subequations}\label{eq:fourier_spacing_condition}%
\begin{align}%
    N_{x} \Delta x \Delta k_{x} &= 2 \pi\\
    N_{y} \Delta y \Delta k_{y} &= 2 \pi
\end{align}%
\end{subequations}%
on the parameters, we can establish orthonormality for the discrete basis exponentials $e^{i\vec{x}_{m,n} \vec{k}_{a,b}^{\vec{\kappa}}}$, in the form
\begin{equation}
 \frac{1}{N_{x}}\frac{1}{N_{y}} \sum_{n=0}^{N_{x}-1} \sum_{m=0}^{N_{y}-1} e^{-i\vec{x}_{m,n} \left( \vec{k}_{a,b}^{\vec{\kappa}} - \vec{k}_{a',b'}^{\vec{\kappa}} \right)} = \delta_{a,a'}\delta_{b,b'}\,.\label{eq:exponential_orthogonality}
\end{equation}
One can see this from the expansion of the exponent
\begin{subequations}\label{eq:exponential_expansion}%
\begin{align}%
 &\vec{x}_{m,n} \left( \vec{k}_{a,b}^{\vec{\kappa}} - \vec{k}_{a',b'}^{\vec{\kappa}}\right) \\
 &\qquad= x_{\textrm{min}}(a-a') \Delta k_{x}+y_{\textrm{min}}(b-b')\, \Delta k_{y}\label{eq:space_offset_phase}\\
 &\qquad+ m (a-a') \frac{2 \pi}{N_x} + n (b-b') \frac{2 \pi}{N_y}\,,\label{eq:2_pi_periodicity}
\end{align}%
\end{subequations}%
with Eq. \eqref{eq:fourier_spacing_condition} substituted. For the situation that  $a \neq a'$, the sum over $m$ over the exponentials of \eqref{eq:2_pi_periodicity} evaluates to zero. A similar statement holds for the sum of the exponentials over $n$ for the case $b \neq b'$. The exponent \eqref{eq:exponential_expansion} turns zero, if we have $a=a'$ and $b=b'$, such that all exponentials in \eqref{eq:exponential_orthogonality} are one and the double sum over $m$ and $n$ in the equation cancels the normalization $(N_x N_y)^{-1}$ and evaluates to one, thus revealing claimed orthonormality of Eq. \eqref{eq:exponential_orthogonality}. Since $N_x N_y$ mutually orthonormal basis exponentials $e^{i\vec{x}_{m,n} \vec{k}_{a,b}^{\vec{\kappa}}}$ exist, the exponentials also form a complete base, where completeness can also be shown explicitly.

On the basis of the shown orthonormality, we can establish the Fourier transform
\begin{subequations}\label{eq:vector_potential_fourier_transform}%
\begin{align}%
 &\vec A'(\vec k^{\vec K}_{a,b},t)\label{eq:vector_potential_position_to_momentum_fourier_transform}\\
 &\quad=\frac{1}{\sqrt{N_x}}\frac{1}{\sqrt{N_y}}\sum_{m=0}^{N_x-1}\sum_{n=0}^{N_y-1}\vec A(\vec x_{m,n},t) e^{-i\vec{x}_{m,n} \vec{k}_{a,b}^{\vec K}}\nonumber
\end{align}%
\end{subequations}%
with inverse%
\begin{subequations}%
\begin{align}%
 &\vec A(\vec x_{m,n},t)\\
 &\quad=\frac{1}{\sqrt{N_x}}\frac{1}{\sqrt{N_y}}\sum_{a=0}^{N_x-1}\sum_{b=0}^{N_y-1}\vec A'(\vec k^{\vec K}_{a,b},t) e^{i\vec{x}_{m,n} \vec{k}_{a,b}^{\vec K}}\nonumber\,,
\end{align}%
\end{subequations}%
which we show for the example of the vector potential with the momentum offset $\vec K$ of the electro-magnetic field.

\subsection{Wave function and operator representations\label{eq:wave_function_representation}}

With the position space representation of a quantum state
\begin{equation}
  \braket{\vec x_{m,n}|\Psi} = \Psi(\vec x_{m,n})
\end{equation}
we can identify the exponentials as position space representation of momentum eigenstates \cite{sakurai2014modern}
\begin{equation}
 \braket{\vec x_{m,n}|\vec{k}_{a,b}^{\vec{\kappa}}} = \frac{1}{\sqrt{N_x}} \frac{1}{\sqrt{N_y}} e^{i\vec{x}_{m,n} \vec{k}_{a,b}^{\vec{\kappa}}}\label{eq:discrete_plane_wave}
\end{equation}
on our grid, such that the orthogonality \eqref{eq:exponential_orthogonality} can be cast into
\begin{subequations}\label{eq:braket_orthogonality}%
\begin{equation}%
 \sum_{n=0}^{N_{x}-1} \sum_{m=0}^{N_{x}-1} \braket{\vec{k}_{a,b}^{\vec{\kappa}}|\vec x_{m,n}} \braket{\vec x_{m,n}|\vec{k}_{a',b'}^{\vec{\kappa}}} = \delta_{a,a'}\delta_{b,b'} \label{eq:position_orthogonality}
\end{equation}%
and likewise%
\begin{equation}%
 \sum_{a=0}^{N_{x}-1} \sum_{b=0}^{N_{x}-1} \braket{\vec x_{m,n}|\vec{k}_{a,b}^{\vec{\kappa}}} \braket{\vec{k}_{a,b}^{\vec{\kappa}}|\vec x_{m',n'}} = \delta_{m,m'}\delta_{n,n'}\,.
\end{equation}%
\end{subequations}%
Eqs. \eqref{eq:braket_orthogonality} and also the completeness of the exponentials $e^{i\vec{x}_{m,n} \vec{k}_{a,b}^{\vec{\kappa}}}$ allow for the identification of the identity operations
\begin{equation}
 \mathds{1}=\sum_{n=0}^{N_{x}-1} \sum_{m=0}^{N_{x}-1} \ket{\vec x_{m,n}}\bra{\vec x_{m,n}}=\sum_{a=0}^{N_{x}-1} \sum_{b=0}^{N_{x}-1} \ket{\vec{k}_{a,b}^{\vec{\kappa}}}\bra{\vec{k}_{a,b}^{\vec{\kappa}}}\,.\label{eq:position_momentum_identity}
\end{equation}
We therefore can implement the inner product between two quantum states $\Psi$ and $\Phi$ as
\begin{equation}
	\braket{ \Psi | \Phi} = \braket{ \Psi | \mathds{1} | \Phi} = \sum_{m=0}^{N_{x}-1}\sum_{n=0}^{N_{y}-1} \Psi(\vec{x}_{m,n})^{\dagger} \Phi(\vec{x}_{m,n})\,,\label{eq:inner_product}
\end{equation}
and the matrix elements of general operators $Q$ with kernel $Q(\vec x_{m,n},\vec x_{m',n'})$ read as
\begin{multline}
 \braket{\Psi|Q|\Phi} = \braket{\Psi|\mathds{1}Q\mathds{1}|\Phi}\\
 =\sum_{m=0 \atop m'=0}^{N_x-1} \sum_{n=0\atop n'=0}^{N_y-1} \Psi(\vec x_{m,n})^\dagger Q(\vec x_{m,n},\vec x_{m',n'}) \Phi(\vec x_{m',n'})\,.
\end{multline}
For the case of an operator which is diagonal in position space
\begin{equation}
 Q(\vec x_{m,n},\vec x_{m',n'}) = Q(\vec x_{m,n}) \delta_{m,m'}\delta_{n,n'} \,,
\end{equation}
like for example the vector potential $\vec A(\vec x,t)$, this matrix element reduces into the 
operator expectation value
\begin{equation}
 \braket{\Psi|Q|\Phi} = \sum_{m=0}^{N_x-1} \sum_{n=0}^{N_y-1} \Psi(\vec x_{m,n})^\dagger Q(\vec x_{m,n}) \Phi(\vec x_{m,n})\,.
\end{equation}
The momentum operator is diagonal in momentum space and reads as
\begin{multline}
 \braket{\vec x_{m,n}|\vec p|\vec x_{m,n}} \\
 = \sum_{a=0}^{N_{x}-1} \sum_{b=0}^{N_{x}-1} \braket{\vec x_{m,n}|\vec{k}_{a,b}^{\vec{\kappa}}} \hbar \vec{k}_{a,b}^{\vec{\kappa}} \braket{\vec{k}_{a,b}^{\vec{\kappa}}|\vec x_{m',n'}}
\end{multline}
in position space, such that for the action at a momentum eigenstate it evaluates as
\begin{multline}%
 \braket{\vec x_{m,n}|\vec p|\vec{k}_{a,b}^{\vec{\kappa}}}
 = \braket{\vec x_{m,n}|\vec p \mathds{1} |\vec{k}_{a,b}^{\vec{\kappa}}} 
 = \hbar \vec{k}_{a,b}^{\vec{\kappa}} \braket{\vec x_{m,n}|\vec{k}_{a,b}^{\vec{\kappa}}}\,,
\end{multline}%
which can be formally deduced from the above relations by inserting the position space identity as indicated. The momentum indices $a$ and $b$ are then fixed by using \eqref{eq:position_orthogonality} and summing over the delta functions in momentum space. Since this holds for every $\vec x_{m,n}$ and the exponentials are orthogonal and complete, we obtain the eigenvalue relation for the momentum operator
\begin{equation}
 \vec p \ket{\vec{k}_{a,b}^{\vec{\kappa}}} = \hbar \vec{k}_{a,b}^{\vec{\kappa}} \ket{\vec{k}_{a,b}^{\vec{\kappa}}}\label{eq:momentum_eigen_value_relation}
\end{equation}
on the grid.

% Position- and momentum space operators are implemented as diagnoal operation in position and momentum space, respectively. In particular 
% If $\hat{\vec x}$ denotes the position operator, it's action at the position space wave function becomes
% \begin{equation}
%  \hat{\vec x} \Phi(\vec{x}_{m,n}) = \vec{x}_{m,n} \Phi(\vec{x}_{m,n})\,.
% \end{equation}
% Correspondingly, the action of the momentum operator $\hat{\vec p}$ is
% \begin{equation}
%  \hat{\vec p} c_{a,b}^{\gamma,s}(t,\vec \kappa) = \hbar \vec{\kappa}_{a,b} c_{a,b}^{\gamma,s}(t,\vec \kappa)\,.
% \end{equation}

\subsection{Relativistic quantum mechanics\label{sec:relativistic_quantum_dynamics}}

We desire to obtain solutions for the time evolution
\begin{equation}
 i \hbar \frac{\partial}{\partial t}\Psi(\vec x,t) = H(t) \Psi(\vec x,t)\,,
\end{equation}
in a relativistic description of quantum mechanics with Hamiltonian
\begin{equation}
 H(t) = H_0 + V(\vec x,t)\,,\label{eq:dirac_hamiltonian}
\end{equation}
where
\begin{equation}
 H_0 = c\, \vec \alpha \cdot \vec p + \beta m c^2 \label{eq:free_hamiltonian}
\end{equation}
is the time-independent, kinetic term and
\begin{equation}
 V(\vec x,t) = - q \, \vec \alpha \cdot \vec A(\vec x,t) + q A_0(\vec x,t)\label{eq:interaction_hamiltonian}
\end{equation}
is the time-dependent interaction of the Dirac equation in canonical form. In this article, we work in Coulomb gauge, with $\vec \nabla \cdot \vec A=0$ and assume the propagation of the electro-magnetic field in a source-free environment, such that the scalar potential vanishes $A_0=0$, see reference \cite{jackson_1999_classical_electro_dyamics}. Additionally, we implement our description in two dimensions, such that $A_3=0$. The Hamiltonian contains the vector $\vec \alpha$ of the the Dirac alpha matrices $\alpha_i$ and the Dirac beta matrix $\beta$ in Dirac representation, see references \cite{ahrens_2020_two_photon_bragg_scattering,ahrens_guan_2022_beam_focus_longitudinal,wachter_2011_relativistic_quantum_mechanics} for details and conventions.

For the perturbative solution in the section below, we require a momentum space formulation in terms of the energy eigensolution bi-spinors
\begin{subequations}\label{eq:bi-spinors}%
\begin{align}%
u_{\vec k}^{+,s} &= \sqrt{\frac{E({\vec k})+m c^2}{2 E({\vec k})}}
\begin{pmatrix}
 \chi^s \\
 \frac{c \vec \sigma \cdot \vec k \hbar}{E(\vec k)+m c^2}\chi^s
\end{pmatrix}\\
u_{\vec k}^{-,s} &= \sqrt{\frac{E(\vec k)+m c^2}{2 E(\vec k)}}
\begin{pmatrix}
 - \frac{c \vec \sigma \cdot \vec k \hbar}{E(\vec k)+m c^2}\chi^s \\
 \chi^s
\end{pmatrix}\,,
\end{align}%
\end{subequations}%
with relativistic energy momentum relation
\begin{equation}
 E(\vec k) = \sqrt{m^2 c^4 + c^2 \hbar^2 \vec k^2}\label{eq:continuous_energy_momentum_relation}
\end{equation}
and vector $\vec \sigma$ of Pauli matrices $\sigma_i$. The spin $s$ can assume the values 0 (spin-up) and 1 (spin-down) with the spinor basis
\begin{equation}
\chi^1 =
\begin{pmatrix}
1 \\ 0
\end{pmatrix}
\,,\qquad
\chi^2 =
\begin{pmatrix}
0 \\ 1
\end{pmatrix}\,.
\end{equation}
We mention that we set the third component of the momentum variable $p_3$ and wave vector $k_3$ to zero in this article, because we are operating on a 2-dimensional grid. This restriction to two dimensions holds especially for the bi-spinor solutions \eqref{eq:bi-spinors}. The bi-spinors \eqref{eq:bi-spinors} fulfill the eigenvalue relation
\begin{equation}\label{eq:energy_eigen_relation}
 H_0(\vec k) u_{\vec k}^{\gamma,s} = \gamma E(\vec k) u_{\vec k}^{\gamma,s}
\end{equation}
for the coefficient matrix
\begin{equation}
 H_0(\vec k) = e^{-i \vec k \cdot \vec x} H_0 e^{i \vec k \cdot \vec x} =  \hbar c\, \vec \alpha \cdot \vec k + \beta m c^2
\end{equation}
of the free Dirac Hamiltonian \eqref{eq:free_hamiltonian}, with $\gamma\in\{+1,-1\}$. They also form a complete, orthonormal base in the four component bi-spinor space, with the identity operation \cite{wachter_2011_relativistic_quantum_mechanics}
\begin{equation}
\sum_{\gamma,s} u_{\vec k}^{\gamma,s}u_{\vec k}^{\gamma,s\,\dagger}\,,
\end{equation}
where the sum iterates over all possible values of $\gamma$ and $s$.

The orthonormality and completeness of the exponentials $e^{i\vec{x}_{m,n} \vec{k}_{a,b}^{\vec{\kappa}}}$, as well as the bi-spinors $u_{\vec k}^{\gamma,s}$ allows us to generalize the exponentials \eqref{eq:discrete_plane_wave} into the bi-spinor basis functions
\begin{equation}
\psi_{a,b}^{\gamma,s}(\vec x_{m,n},\vec \kappa) = \braket{\vec x_{m,n}|\vec{k}_{a,b}^{\vec{\kappa}}} u_{\vec k^{\vec \kappa}_{a,b}}^{\gamma,s}\,.\label{eq:bi_spinor_plane_wave_basis_states}
\end{equation}
We combine the indices $\gamma$, $s$, $a$ and $b$ in the index set $I=\{\gamma,s,a,b\}$ and introduce the ket-association for the plane wave bi-spinor
\begin{equation}
 \ket{\psi_I^{\vec \kappa}} = \ket{\vec k^{\vec \kappa}_{a,b}} u_{\vec k^{\vec \kappa}_{a,b}}^{\gamma,s} \label{eq:bi_spinor_plane_wave_basis_kets}
\end{equation}
on our two-dimensional grid space. Inserting the basis states \eqref{eq:bi_spinor_plane_wave_basis_kets} into the inner product \eqref{eq:inner_product} establishes the bra-ket orthonormality
\begin{equation}
	\braket{\psi_{I'}^{\vec{\kappa}} \mid \psi_{I}^{\vec{\kappa}} } = \delta_{I',I}\,,
\end{equation}
where we have introduced the the index set Kronecker delta
\begin{equation}
  \delta_{I',I} = \delta_{\gamma',\gamma}\delta_{s',s}\delta_{a',a}\delta_{b',b}\,.
\end{equation}
From the completeness of the exponentials $e^{i\vec{x}_{m,n} \vec{k}_{a,b}^{\vec{\kappa}}}$ and the bi-spinors $u_{\vec k}^{\gamma,s}$, we can also extend the notion for the identity operation \eqref{eq:position_momentum_identity} into
\begin{equation}
 \mathds{1}=\sum_I \ket{\psi_I^{\vec \kappa}}\bra{\psi_I^{\vec \kappa}}\,.\label{eq:bi_spinor_space_identity}
\end{equation}
From the eigenvalue relation of the momentum operator \eqref{eq:momentum_eigen_value_relation} and the eigenvalue relation of the bi-spinors \eqref{eq:energy_eigen_relation}, we also conclude the eigenvalue relation
\begin{equation}
 H_0 \ket{\psi_I^{\vec \kappa}} = H_0(\vec k^{\vec \kappa}_{a,b}) \ket{\vec k^{\vec \kappa}_{a,b}} u_{\vec k^{\vec \kappa}_{a,b}}^{\gamma,s} = \gamma E_{a,b}^{\vec \kappa} \ket{\psi_I^{\vec \kappa}}\label{eq:energy_eigenvalue_relation_dirac}
\end{equation}
for the bi-spinor basis functions $\ket{\psi_I^{\vec \kappa}}$ with respect to the free Hamiltonian, where we use the notion
\begin{equation}
 E\left(\vec k^{\vec \kappa}_{a,b}\right) = E_{a,b}^{\vec \kappa}
\end{equation}
for the energy \eqref{eq:continuous_energy_momentum_relation} on grid space. The energy eigenvalue relation \eqref{eq:energy_eigenvalue_relation_dirac} for the bi-spinor plane waves $\ket{\psi_I^{\vec \kappa}}$ allows for the introduction of the wave function in the interaction picture
\begin{equation}
 \ket{\tilde \Psi(t)}_I = e^{i H_0 t/\hbar} \ket{\Psi(t)}\,.
\end{equation}

With these conventions, the wave function can be expanded in terms of the basis states as
\begin{subequations}%
\begin{equation}%
 \ket{\tilde \Psi(t)}=\sum_I c_I^{\vec \kappa}(t)\ket{\psi_I^{\vec \kappa}}\,,\label{eq:psi_momentum_to_position_fourier_transform}
\end{equation}%
with the momentum space expansion coefficients $c_I^{\vec \kappa}(t) \ \Leftrightarrow \ c_{a,b}^{\gamma,s}(t,\vec \kappa)$ in the interaction picture, given by%
\begin{equation}%
 c_I^{\vec \kappa}(t) = \braket{\psi_I^{\vec \kappa}|\tilde \Psi(t)}\,.\label{eq:psi_position_to_momentum_fourier_transform}
\end{equation}%
\end{subequations}%
Note, that Eq. \eqref{eq:psi_momentum_to_position_fourier_transform} constitutes a Fourier transformation of the quantum state $\ket{\tilde \Psi(t)}$, with inverse \eqref{eq:psi_position_to_momentum_fourier_transform} and can be cast into an explicit component notion, similarly to Eqs. \eqref{eq:vector_potential_fourier_transform} for the vector potential.

\subsection{Introduction of the external Gaussian beam potential\label{sec:gaussian_beam_potentials}}

Having established notions for the electron wave function, the vector potential and relativistic quantum equation of motion, we now want to introduce the potential of the external Gaussian beam for the investigation the electron dynamics in a focused standing wave laser beam. In context of the references \cite{ahrens_guan_2022_beam_focus_longitudinal,Quesnel_1998_gaussian_beam_coulomb_gauge}, we denote the potentials
\begin{subequations}\label{eq:gaussian_beam_potentials}%
\begin{align}%
A_{x,d}(\vec x,t)=&-2dA_{0}\frac{w_{0}}{w}\epsilon\frac{y}{w}\exp\left(-\frac{y^2}{w^2}\right)\cos\left(\phi_{G,d}^{(1)}\right)\eta(t)\\
A_{y,d}(\vec x,t)=&-A_{0}\frac{w_{0}}{w}\exp\left(-\frac{y^2}{w^2}\right)\sin\left(\phi_{G,d}\right)\eta(t)\,.
\end{align}%
\end{subequations}%
with the phases
\begin{subequations}\label{eq:gaussian_beam_phases}%
\begin{align}%
\phi_{G,d}(\vec x,t)=&\omega t-dk_{L}x+\tan^{-1}\left(\frac{dx}{x_{R}}\right)-\frac{d x y^2}{x_{R}w^^2}\\
\phi_{G,d}^{(1)}(\vec x,t)=&\phi_{G,d}+\tan^{-1}\left(\frac{dx}{x_{R}}\right)
\end{align}%
\end{subequations}%
and temporal envelope
\begin{equation}\label{eq:field_temporal_envelope}
 \eta(t) =
\begin{cases}
 0 & \text{, for } t<0\\
\sin^2(\Omega t) & \text{, for } \ 0\le t\le T\\
 0 & \text{, for }\ T<t\,.
\end{cases}
\end{equation}
The potentials in Eqs. \eqref{eq:gaussian_beam_potentials}-\eqref{eq:field_temporal_envelope} are representing a Gaussian beam which is propagating along the $x$-axis, with linear polarization along the $y$-axis. Parameters are the potential amplitude $A_0$, wave number $k_L=2 \pi/\lambda$ (with wave length $\lambda$) and beam waist of the Gaussian beam $w_0$. The diffraction angle $\epsilon$ is related to the introduced quantities by $\epsilon=1/(k_L w_0)$, in the convention of the originating reference \cite{Quesnel_1998_gaussian_beam_coulomb_gauge}, and $x_R$ is the Rayleigh length $x_R=k_L w_0^2/2$.  Furthermore, the variable $w$ is representing the $x$-dependent waist of the beam
\begin{equation}
w(x)=w_{0}\sqrt{1+\frac{x^2}{x_{R}^2}}\,.
\end{equation}
A $\sin^2$ temporal envelope $\eta(t)$ in Eq. \eqref{eq:field_temporal_envelope} is modeling a smooth ramp-up and ramp-down of the external field, in accordance with similar procedures in references \cite{ahrens_bauke_2013_relativistic_KDE,ahrens_2017_spin_filter,ahrens_2012_phdthesis_KDE}, where $T$ is the interaction duration and we set $\Omega=\pi/T$. The index $d \in \{-1,+1\}$ in Eqs. \eqref{eq:gaussian_beam_potentials} and \eqref{eq:gaussian_beam_phases} specifies the propagation direction of the Gaussian beam.

For using the potentials \eqref{eq:gaussian_beam_potentials} in our perturbative solution approach below, we further expand the trigonometric functions into exponentials of the form
\begin{subequations}\label{eq:gaussian_beam_exponentials}%
\begin{align}%
A_{x,d,o}(\vec x,t)=&-dA_{0}\frac{w_{0}}{w}\epsilon\frac{y}{w}\exp\left(-\frac{y^2}{w^2}\right)\nonumber\\
&\qquad\qquad\times\exp\left(i \,o\, \phi_{G,d}^{(1)}\right)\eta(t)\label{eq:gaussian_beam_exponentials_longitudinal}\\
A_{y,d,o}(\vec x,t)=&i o \frac{A_{0}}{2}\frac{w_{0}}{w}\exp\left(-\frac{y^2}{w^2}\right)\nonumber\\
&\qquad\qquad\times\exp\left(i \,o\, \phi_{G,d}\right)\eta(t)\,,\label{eq:gaussian_beam_exponentials_transverse}
\end{align}%
\end{subequations}%
where the index $o\in\{-1,+1\}$ can be associated with photon emission or absorption between laser field and electron \cite{ahrens_2020_two_photon_bragg_scattering}. Accordingly, in a similar fashion as in reference \cite{ahrens_guan_2022_beam_focus_longitudinal}, we can write
\begin{equation}
 A_{j,d}(\vec x,t) = \sum_o A_{j,d,o}(\vec x,t)\,.\label{eq:gaussian_exponential_expansion_sum}
\end{equation}
\begin{figure}%
\includegraphics[width=0.5 \textwidth]{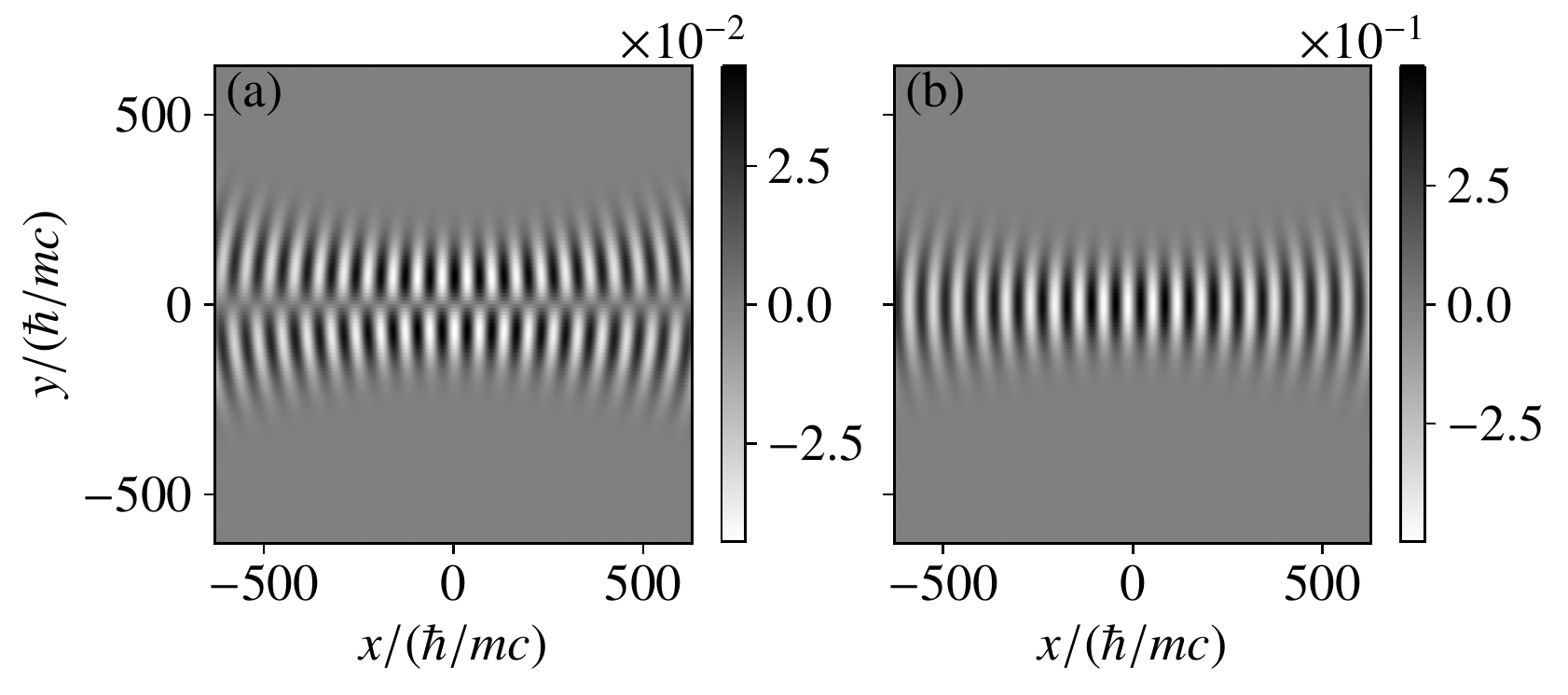}
\caption{\label{fig:vector_potential_position_space_high_resolution} Position space illustration of the Gaussian beam exponentials \eqref{eq:gaussian_beam_exponentials}. We display Eq. \eqref{eq:gaussian_beam_exponentials_longitudinal} (left panel) and Eq. \eqref{eq:gaussian_beam_exponentials_transverse} (right panel) with the beam parameters $k_L=0.1\,mc/\hbar$, $\epsilon=0.1$ and $A_0=1mc^2/q$ for $d=1$ and $o=-1$ at a grid resolution of $N_x=1024$, $N_y=128$, as discussed in section \ref{sec:gaussian_beam_potentials}. The real part of the beams \eqref{eq:gaussian_beam_exponentials} is shown, for illustrating the wave-like structure of the Gaussian beam.}%
\end{figure}%
For elucidating the role of the interaction potential in our perturbative solution method, we display the real part of Eqs. \eqref{eq:gaussian_beam_exponentials} in Fig. \ref{fig:vector_potential_position_space_high_resolution}. The parameters used in Fig. 2 are $k_L=0.1\,mc/\hbar$, $\epsilon=0.1$ and $A_0=1mc^2/q$. The simulation area has an extension of
\begin{subequations}%
\begin{align}%
 x_\textrm{min}&=-20 \lambda\\
 y_\textrm{min}&=-20 \lambda
\end{align}%
\end{subequations}%
for the minimum axis values and
\begin{subequations}%
\begin{align}%
 x_\textrm{max}&=x_{N_x-1}=20 \lambda\\
 y_\textrm{max}&=y_{N_y-1}=20 \lambda
\end{align}%
\end{subequations}%
for the maximum axis values. For the grid resolution we set $N_x=1024$ along the $x$-direction and $N_y=128$ along the $y$-direction. With the simulation box width
\begin{subequations}%
\begin{align}%
 x_\textrm{w}&=x_\textrm{max} - x_\textrm{min} \\
 y_\textrm{w}&=y_\textrm{max} - y_\textrm{min}
\end{align}%
\end{subequations}%
this implies the grid spacing
\begin{subequations}%
\begin{align}%
 \Delta x&=\frac{x_\textrm{w}}{N_x - 1} \\
 \Delta y&=\frac{y_\textrm{w}}{N_y - 1}\,.
\end{align}%
\end{subequations}%
In a similar manner, the simulation box width in momentum space is given by
\begin{subequations}\label{eq:momentum_space_width}%
\begin{align}%
 k_{x,\textrm{w}}&=\Delta k_x(N_x - 1)\\
 k_{y,\textrm{w}}&=\Delta k_y(N_y - 1)\,,
\end{align}%
\end{subequations}%
where the momentum grid spacings $\Delta k_x$ and $\Delta k_y$ are necessarily implied by condition \eqref{eq:fourier_spacing_condition}. We now want to show the momentum space distribution of the Gaussian beam vector potential. For illustration purposes we first choose a symmetric frame for the coordinates in momentum space with
\begin{subequations}\label{eq:momentum_space_lower_bound_high_resolution}%
\begin{align}%
  K_x&=-\frac{k_{x,\textrm{w}}}{2}\label{eq:momentum_space_lower_bound_high_resolution_x}\\
  K_y&=-\frac{k_{y,\textrm{w}}}{2}\,,
\end{align}%
\end{subequations}%
such that we obtain for the maximum coordinates
\begin{subequations}\label{eq:momentum_space_upper_bound_high_resolution}%
\begin{align}%
 K_{x,\textrm{max}}&=k_{x,N_x-1}=\frac{k_{x,\textrm{w}}}{2}\label{eq:momentum_space_upper_bound_high_resolution_x}\\
 K_{y,\textrm{max}}&=k_{y,N_y-1}=\frac{k_{y,\textrm{w}}}{2}\,.
\end{align}%
\end{subequations}%

\begin{figure}
\includegraphics[width=0.5 \textwidth]{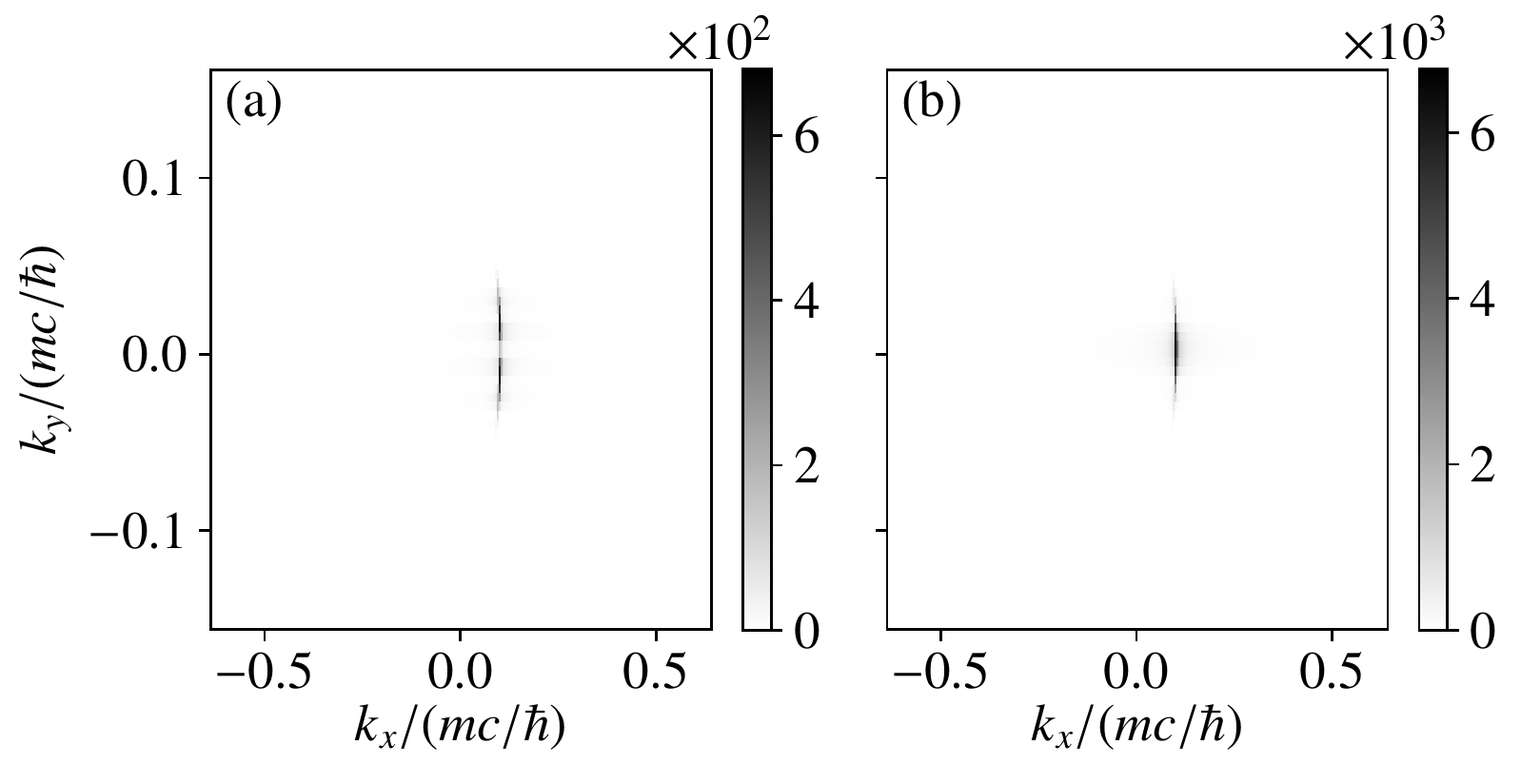}
\caption{\label{fig:vector_potential_momentum_space_high_resolution} The Gaussian beam vector potential in momentum space. The Fourier transform \eqref{eq:vector_potential_position_to_momentum_fourier_transform} of the potentials in Fig. \ref{fig:vector_potential_position_space_high_resolution} is shown, with the identical parameters. We display the absolute value of the Fourier transform, but use the complex input fields of the Gaussian beam exponentials \eqref{eq:gaussian_beam_exponentials} for the transform, in contrast to the real part which is displayed in Fig. \ref{fig:vector_potential_position_space_high_resolution}. Note, that we are showing a sub-region of momentum space, smaller than the minimum value $\vec K$ and maximum value $\vec K_\textrm{max}$ of Eqs. \eqref{eq:momentum_space_lower_bound_high_resolution} and \eqref{eq:momentum_space_upper_bound_high_resolution}, for better visibility.}
\end{figure}

With the mentioned parameters we are able to perform the Fourier transform \eqref{eq:vector_potential_position_to_momentum_fourier_transform} of the Gaussian beam exponentials \eqref{eq:gaussian_beam_exponentials} as presented in Fig. \ref{fig:vector_potential_momentum_space_high_resolution}. We see a double peak structure in Fig. \ref{fig:vector_potential_momentum_space_high_resolution}(a) and a single peak in Fig. \ref{fig:vector_potential_momentum_space_high_resolution}(b). The reason for the double peak structure is discussed in reference \cite{ahrens_guan_2022_beam_focus_longitudinal}, if one accounts for the absolute value of the Fourier transform being displayed in Fig. \ref{fig:vector_potential_momentum_space_high_resolution}(a). More importantly, we recognize that both peaks in the two panels of Fig. \ref{fig:vector_potential_momentum_space_high_resolution} are centered at the position $k_L \vec e_x$ in momentum space, which is caused by the exponential $e^{i k_L x}$ in the combined equations \eqref{eq:gaussian_beam_phases} and \eqref{eq:gaussian_beam_exponentials} for $o=1$ and $d=-1$. Consequently, if the product $od$ turns positive (+1), the peak appears mirror-symmetrically at position $- k_L \vec e_x$ in momentum space. In terms of the interpretation of an action in physics, the peaks at $\pm k_L \vec e_x$ are imposing a momentum transfer of momentum $\pm \hbar k_L \vec e_x$ at the electron wave function.

\begin{figure}
	\includegraphics[width=0.5 \textwidth]{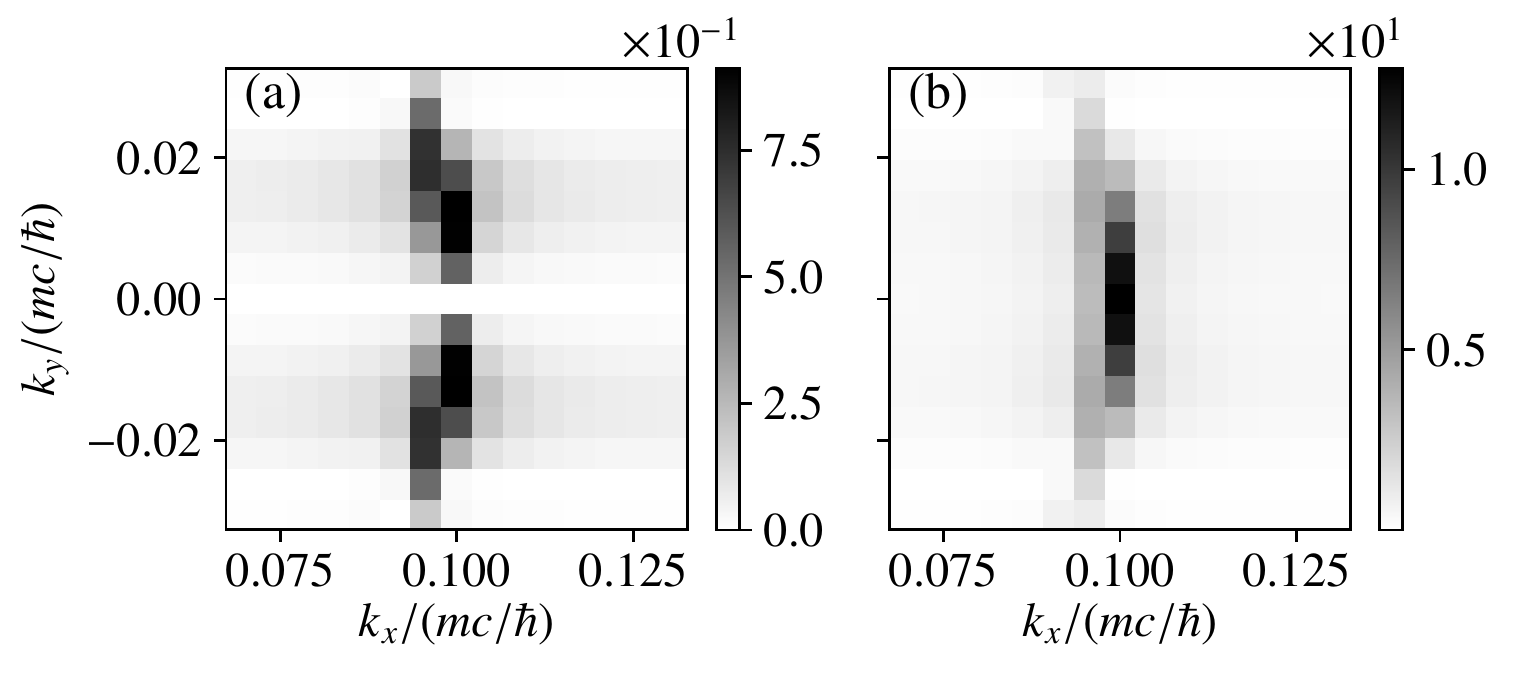}\caption{\label{fig:vector_field_momentum_space_magnified} Resolution adjusted Gaussian beam vector potential in momentum space. We repeat the Fourier transform in Fig. \ref{fig:vector_potential_momentum_space_high_resolution}, but now with grid resolution $N_x=15$ and $N_y=15$, and the offset momentum $\vec K$ adjusted as in Eq. \eqref{eq:momentum_space_lower_bound}. In contrast to Fig. \ref{fig:vector_potential_momentum_space_high_resolution}, we display the whole momentum grid space in this figure.}
\end{figure}

As mentioned in the introduction and explained in the coming section about the perturbative formalism, we want to simulate the two-photon diffraction of the Kapitza-Dirac effect in the Bragg regime, for which the electron momentum changes from $-\hbar k_L$ to $\hbar k_L$ along the $x$-axis, by the transfer of two photons with momentum $\hbar k_L$. This approach is consistent with the procedures in references \cite{ahrens_bauke_2013_relativistic_KDE,ahrens_2017_spin_filter,ahrens_2020_two_photon_bragg_scattering,ahrens_guan_2022_beam_focus_longitudinal}. Furthermore, as discussed in section \ref{sec:execution_scaling}, the runtime of our method scales quadratically with the grid resolution of each dimension. We therefore want to resolve the momentum peak in momentum space at a desirable low grid resolution. Therefore, we center the momentum space grid around the peak at $k_L \vec e_x$, by setting
\begin{subequations}\label{eq:momentum_space_lower_bound}%
\begin{align}%
  K_x&=-\frac{k_{x,\textrm{w}}}{2} + k_L\\
  K_y&=-\frac{k_{y,\textrm{w}}}{2}
\end{align}%
\end{subequations}%
with the implication
\begin{subequations}\label{eq:momentum_space_upper_bound}%
\begin{align}%
  K_{x,\textrm{max}}&=\frac{k_{x,\textrm{w}}}{2} + k_L\\
  K_{y,\textrm{max}}&=\frac{k_{y,\textrm{w}}}{2}
\end{align}%
\end{subequations}%
instead of Eqs. \eqref{eq:momentum_space_lower_bound_high_resolution} and \eqref{eq:momentum_space_upper_bound_high_resolution} and reduce the grid resolution to $N_x=15$ and $N_y=15$, resulting in the Fourier transform displayed in Fig. \ref{fig:vector_field_momentum_space_magnified}.

\subsection{Time-dependent perturbation theory\label{sec:perturbation_theory}}

We have introduced the Dirac Hamiltonian \eqref{eq:dirac_hamiltonian} such that it is split up into the time-independent kinetic part $H_0$ \eqref{eq:free_hamiltonian} and the time-dependent interaction part from the potentials $V(\vec x,t)$ \eqref{eq:interaction_hamiltonian}, which is the requirement for applying time-dependent perturbation theory \cite{sakurai2014modern}. In our article, we aspire the theoretical description of a two-photon interaction in the Bragg regime of the Kapitza-Dirac effect \cite{batelaan_2000_KDE_first,batelaan_2007_RMP_KDE}, demanding for second order time dependent perturbation theory, with one order in perturbation theory for each photon interaction, see references \cite{ahrens_bauke_2012_spin-kde,ahrens_bauke_2013_relativistic_KDE,ahrens_2012_phdthesis_KDE} for details. The time-evolution operator of second order time-dependent perturbation theory is given by \cite{sakurai2014modern}
\begin{equation}
U(t,t_{0})=\left(-\frac{i}{\hbar}\right)^{2}\int_{t_0}^{t} dt_2 \int_{t_0}^{t_2} dt_1 V_{\textrm{int}}(t_2)V_{\textrm{int}}(t_1)\,,\label{eq:second_order_time_dependent_perturbation_theory}
\end{equation}
with the interaction Hamiltonian in the interaction picture
\begin{equation}
 V_\textrm{int} = e^{i H_0 t/\hbar} V(t) e^{-i H_0 t/\hbar}\,.
\end{equation}
For computing the action of the time-evolution operator $U$ at energy eigenstates $\ket{\psi_I^{\vec \kappa}}$, we first consider the action of the operator product $V_{\textrm{int}}(t_2)V_{\textrm{int}}(t_1)$ at $\ket{\psi_I^{\vec \kappa}}$. By inserting the identity operator \eqref{eq:bi_spinor_space_identity}, we can write
\begin{multline}\label{eq:interaction_operators_application}
 V_{\textrm{int}}(t_2)V_{\textrm{int}}(t_1)\ket{\psi_I^{\vec \kappa}}
 = \sum_{I''} \sum_{I'} \Bigg[\Ket{\psi_{I''}^{\vec \kappa''}}\\\times\Braket{\psi_{I''}^{\vec \kappa''}|V_{\textrm{int}}(t_2)|\psi_{I'}^{\vec \kappa'}}\Braket{\psi_{I'}^{\vec \kappa'}|V_{\textrm{int}}(t_1)|\psi_I^{\vec \kappa}}\Bigg]\,,
\end{multline}
with the sums running over each possible value of the indices $I'$ and $I''$, where primed and double primed index sets $I'$ and $I''$ also contain primed and double primed variables $\{\gamma',s',a',b'\}$, $\{\gamma'',s'',a'',b''\}$, respectively. The momentum coordinates of the primed and double primed index sets are also implemented with different momentum offsets $\vec \kappa'$ and $\vec \kappa''$.

The operator application in Eq. \eqref{eq:interaction_operators_application} can further be rewritten by working out the matrix elements $\braket{\Psi_{I'}^{\vec{\kappa} '} | V_\textrm{int}(t) | \Psi_{I}^{\vec{\kappa}}}$. We obtain
\begin{widetext}
\begin{multline}\label{eq:interaction_matrix_element}
 \Braket{\Psi_{I'}^{\vec{\kappa} '} | V_\textrm{int}(t) | \Psi_{I}^{\vec{\kappa}}}
 = \Braket{\Psi_{I'}^{\vec{\kappa} '} | e^{i H_0 t/\hbar} V(t) e^{-i H_0 t/\hbar} | \Psi_{I}^{\vec{\kappa}}} 
  = \Braket{\Psi_{I'}^{\vec{\kappa} '} | V(t) | \Psi_{I}^{\vec{\kappa}}} e^{i(\gamma' E_{a',b'}^{{\vec \kappa}'}-\gamma E_{a,b}^{\vec \kappa})t/\hbar} \\
 = \sum_{m=0}^{N_{x}-1} \sum_{n=0}^{N_{y}-1} \psi_{a',b'}^{ \gamma', s'}(\vec{x}_{m,n}, \vec{\kappa}' )^{\dagger} \left[ -q \,\vec{\alpha} \cdot \vec{A} (\vec{x}_{m,n},t) \right] \psi_{a,b}^{ \gamma, s}(\vec{x}_{m,n}, \vec{\kappa} ) e^{i(\gamma' E_{a',b'}^{{\vec \kappa}'}-\gamma E_{a,b}^{\vec \kappa})t/\hbar} \\
 =  -q \frac{1}{N_{x}}\frac{1}{N_{y}} \sum_{m=0}^{N_{x}-1} \sum_{n=0}^{N_{y}-1} \sum_{j=1}^{3} {u_{\vec{k}_{a',b'}^{\vec{\kappa}'}}^{ \gamma', s'}}^{\dagger} \alpha_{j} u_{\vec{k}_{a,b}^{\vec{\kappa}'}}^{ \gamma, s} A_{j}(\vec{x}_{m,n},t) e^{-i \vec{x}_{m,n}( \vec{k}_{a',b'}^{\vec{\kappa}'} - \vec{k}_{a,b}^{\vec{\kappa}})} e^{i(\gamma' E_{a',b'}^{{\vec \kappa}'}-\gamma E_{a,b}^{\vec \kappa})t/\hbar}\,.
\end{multline}
\end{widetext}
By identifying the momentum offset of the vector potential as $\vec K = \vec \kappa' - \vec \kappa$ and setting
\begin{subequations}\label{eq:interaction_matrix_element_indices}%
\begin{align}%
 a_1 &= a'  - a  & b_1 &= b'  - b \\
 a_2 &= a'' - a' & b_2 &= b'' - b'
\end{align}%
\end{subequations}%
we can write
\begin{equation}\label{eq:interaction_offset_shift}
	\vec{k}_{a',b'}^{\vec{\kappa}'} - \vec{k}_{a,b}^{\vec{\kappa}}=\vec{\mathbf{K}} + a_1 \, \Delta k_{x}\, \vec{e}_{x} + b_1 \, \Delta k_{y} \, \vec{e}_{y} = \vec k_{a_1,b_1}^{\vec{\mathbf{K}}}
\end{equation}
for the first interaction at $t_1$. A similar notion can also be written for the case of the second interaction at $t_2$. Eq. \eqref{eq:interaction_offset_shift} allows us to identify and substitute the Fourier transform \eqref{eq:vector_potential_position_to_momentum_fourier_transform} in the matrix element \eqref{eq:interaction_matrix_element}, resulting in
\begin{multline}\label{eq:interaction_matrix_element_computed}
\Braket{ \Psi_{I'}^{\vec{\kappa}'}  | V_{\textrm{int}}(t) | \Psi_{I}^{\vec{\kappa}} } \\
= -q \frac{1}{\sqrt{N_{x}}}\frac{1}{\sqrt{N_{y}}} \sum_{j=1}^{3} {u_{\vec{k}_{a',b'}^{\vec{\kappa}'}}^{\gamma',s'}}^{\dagger} \alpha_{j} u_{\vec{k}_{a,b}^{\vec{\kappa}}}^{\gamma,s} \\
\times A_{j}'(\vec k_{a'-a,b'-b}^{\vec{\mathbf{K}}},t)e^{i(\gamma' E_{a',b'}^{{\vec \kappa}'}-\gamma E_{a,b}^{\vec \kappa})t/\hbar}\,.
\end{multline}
We see that the momentum offsets $\vec \kappa'$ and $\vec \kappa''$ are created by the action of the interaction at the initial state with momentum offset $\vec \kappa$. The spacing between $\vec \kappa$ and $\vec \kappa'$, as well as the spacing between $\vec \kappa'$ and $\vec \kappa''$ is mediated by the momentum offset $\vec K$ of the interaction.

\begin{figure}
\includegraphics[width=0.48 \textwidth]{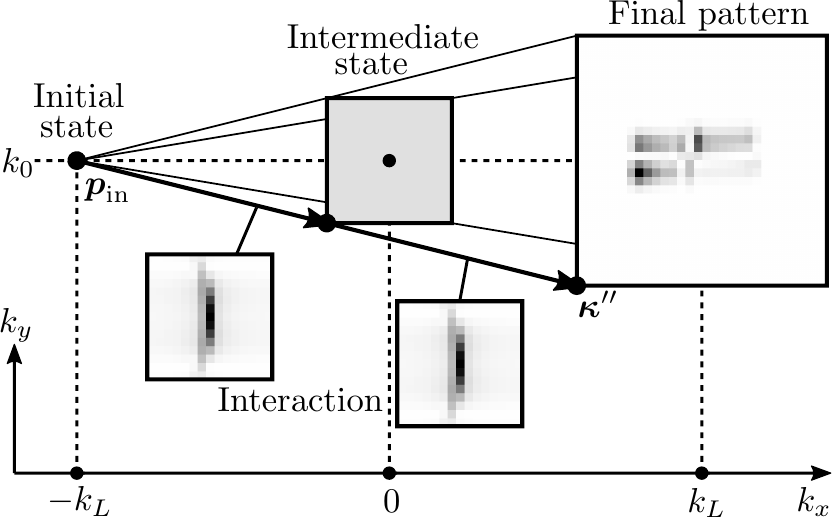}% Here is how to import EPS art
\caption{\label{fig:perturbative_illustration}Illustration of the scattering dynamics in momentum space. In terms interpreting Eq. \eqref{eq:interaction_operators_application}, the initial quantum state with momentum $\vec p_{\textrm{in}}$ is mapped twice by the vector potential of the Gaussian laser beam in the form of $V_\textrm{int}(t_1)$ and $V_\textrm{int}(t_2)$, as represented by the miniaturized reproduction of Fig. \ref{fig:vector_field_momentum_space_magnified}(b). By iterating over the coordinate indices $a_1, a_2, b_1, b_2$ of the potential, the grid geometry of non-vanishing entries for the quantum state in momentum space evolves into $(N_x-1)\times(N_y-1)$ entries for the intermediate state and $2(N_x-1)\times2(N_y-1)$ for the final state, as explicitly denoted in Eqs. \eqref{eq:intermediate_momentum_space_grid}-\eqref{eq:final_momentum_space_grid}. Thereby, the electron momentum is effectively diffracted from momentum $-\hbar k_L \vec e_x + \hbar k_0 \vec e_y$ to approximately $\hbar k_L \vec e_x + \hbar k_0 \vec e_y$.}
\end{figure}

The expansion into multiple states in Eq. \eqref{eq:interaction_operators_application} allows for an instructive explanation of our method, which we have sketched in Fig. \ref{fig:perturbative_illustration}. The initial state for our calculation is a momentum eigenstate with momentum
\begin{equation}
\vec p_{\textrm{in}}=-k_L \vec e_x + k_0 \vec e_y\,,\label{eq:initial_momentum}
\end{equation}
where $k_L$ is the laser wave number and $k_0$ is the electron transverse momentum wave number. In previous calculations \cite{ahrens_bauke_2013_relativistic_KDE,ahrens_2017_spin_non_conservation,ahrens_2020_two_photon_bragg_scattering} the plane wave laser field was transferring a momentum $\hbar k_L \vec e_x$, such that after two photon interactions, the final state at $k_L \vec e_x + k_0 \vec e_y$ is reached, via an intermediate state at $k_0 \vec e_y$. In the present description, we no longer use a plane wave field, resulting in a continuum of momenta centrally arranged around the momentum $\hbar k_L \vec e_x$, as illustrated by the two ``Interaction'' panels of a shrunken picture of Fig. \ref{fig:vector_field_momentum_space_magnified}(b), which are displayed in Fig. \ref{fig:perturbative_illustration}. Therefore, the final pattern in Fig. \ref{fig:perturbative_illustration} (shrunken picture of Fig. \ref{fig:final_diffraction_pattern}) also results in a quasi continuum of momentum eigenstates. The method is an extension of our attempt in reference \cite{ahrens_guan_2022_beam_focus_longitudinal}, where we considered the additional momentum transfer of $k_z$ in the transverse direction (with respect to the laser beam). This allowed for a discrete number of momentum eigenstates, which were written down explicitly in \cite{ahrens_guan_2022_beam_focus_longitudinal}, and are contrasted by the continuum of momentum eigenstates in our present approach. In this context, Fig. \ref{fig:perturbative_illustration} in this article is an extension of the interaction dynamics which was sketched in Fig. 1 of reference \cite{ahrens_guan_2022_beam_focus_longitudinal}.

According to this interaction we consider the grid layout for the electron as in the following. The initial momentum eigenstate with momentum $\vec p_\textrm{in}=\vec \kappa$ is initialized on a single grid point with coordinates $a=0$ and $b=0$. From the first application of the interaction \eqref{eq:interaction_matrix_element_computed}, the intermediate state becomes the coordinates as in Eq. \eqref{eq:momentum_space_grid}
\begin{subequations}\label{eq:intermediate_momentum_space_grid}%
\begin{align}%
k_{x,a'}&=\kappa'_x+a'\,\Delta k_x \,,\quad a' \in \{ 0,\dots,N_{x}-1  \}\\
k_{y,b'}&=\kappa'_y+b'\,\Delta k_y \,,\quad b' \in \{ 0,\dots,N_{y}-1  \}\,,
\end{align}%
\end{subequations}%
with
\begin{subequations}%
\begin{align}%
  \kappa'_x&=-\frac{k_{x,\textrm{w}}}{2}\\
  \kappa'_y&=-\frac{k_{y,\textrm{w}}}{2} + k_0
\end{align}%
\end{subequations}%
and the maximum momentum
\begin{subequations}%
\begin{align}%
  k'_{x,\textrm{max}}&=\frac{k_{x,\textrm{w}}}{2}\\
  k'_{y,\textrm{max}}&=\frac{k_{y,\textrm{w}}}{2} + k_0\,.
\end{align}%
\end{subequations}%
On the second application of the interaction \eqref{eq:interaction_matrix_element_computed} this coordinate layout then extends for the final state into
\begin{subequations}\label{eq:final_momentum_space_grid}%
\begin{align}%
k_{x,a''}&=\kappa''_x+a''\,\Delta k_x \,,\quad a'' \in \{ 0,\dots,2(N_{x}-1)  \}\\
k_{y,b''}&=\kappa''_y+b''\,\Delta k_y \,,\quad b'' \in \{ 0,\dots,2(N_{y}-1)  \}\,,
\end{align}%
\end{subequations}%
including the offset
\begin{subequations}%
\begin{align}%
  \kappa''_x&=-k_{x,\textrm{w}} + k_L\\
  \kappa''_y&=-k_{y,\textrm{w}} + k_0
\end{align}%
\end{subequations}%
and the maximum value of the final coordinates
\begin{subequations}%
\begin{align}%
  k''_{x,\textrm{max}}&=k_{x,\textrm{w}} + k_L\\
  k''_{y,\textrm{max}}&=k_{y,\textrm{w}} + k_0\,.
\end{align}%
\end{subequations}%

\section{Implementation details\label{sec:implementation_details}}

\subsection{Time-propagation and interaction matrix elements\label{sec:time_propagation}}

For denoting our implementation for the perturbative time-evolution operator \eqref{eq:second_order_time_dependent_perturbation_theory}, we write its action at the interaction picture wave function in terms of the projection
\begin{subequations}\label{eq:general_time_evolution}%
\begin{align}%
 c_{I''}^{\vec \kappa''}(T)
 &= \Braket{\psi_{I''}^{\vec \kappa''}|\tilde \Psi(T)}
 = \Braket{\psi_{I''}^{\vec \kappa''}|U(T,0)|\tilde \Psi(0)}\\
 &= \sum_{I}\Braket{\psi_{I''}^{\vec \kappa''}|U(T,0)|\psi_{I}^{\vec \kappa}}\Braket{\psi_{I}^{\vec \kappa}|\tilde \Psi(0)}\\
 &= \sum_{I} U_{I'',I}^{\vec \kappa'',\vec \kappa}(T,0) c_{I}^{\vec \kappa}(0)\,,
\end{align}%
\end{subequations}%
where we see that the expansion coefficients of the initial wave function $c_{I}^{\vec \kappa}(0)$ at time $0$ are propagated to the expansion coefficients $c_{I''}^{\vec \kappa''}(T)$ of the final wave function at time $T$ by the propagation matrix
\begin{subequations}\label{eq:propagator_matrix_element}%
\begin{align}%
 U_{I'',I}^{\vec \kappa'',\vec \kappa}(T,0)&=\Braket{\psi_{I''}^{\vec \kappa''}|U(T,0)|\psi_I^{\vec \kappa}} \\
 &\qquad\Leftrightarrow \ U_{a'',b'';a,b}^{\gamma'',s'';\gamma,s}(T,0,\vec \kappa'',\vec \kappa)\,.\label{eq:explicit_propagator_matrix_element}
\end{align}%
\end{subequations}%
Within this work, we obtain the value of the matrix elements \eqref{eq:propagator_matrix_element} from the perturbative expression \eqref{eq:second_order_time_dependent_perturbation_theory}. The interaction matrix elements \eqref{eq:interaction_matrix_element_computed}, which are inevitably implied from the operator application \eqref{eq:interaction_operators_application} only have physically valid entries for momenta in the range \eqref{eq:momentum_space_grid}, with momentum offset $\vec K$ and interaction matrix index variables as in Eq. \eqref{eq:interaction_matrix_element_indices}. Thereby, the grid indices $a',a'',b',b''$ of the index sets $I'$ and $I''$ in Eq. \eqref{eq:interaction_operators_application} are implied by the indices of the interactions $a_1,a_2,b_1,b_2$, in \eqref{eq:interaction_matrix_element_indices} and assume the specific form as in Eqs. \eqref{eq:intermediate_momentum_space_grid}-\eqref{eq:final_momentum_space_grid}.

\subsection{Laser field contributions of the two-photon Kapitza-Dirac effect and time integration in time-dependent perturbation theory\label{sec:laser_field_time_integration}}

Regarding the external field, we encounter the additional property that the standing light wave, which is assumed for Kapitza-Dirac scattering, is an equal amplitude superposition of left and right traveling wave, in which Eq. \eqref{eq:gaussian_exponential_expansion_sum} is summed up as
 \begin{equation}
 A_{j}(\vec x,t) = \sum_d A_{j,d}(\vec x,t) = \sum_{d,o} A_{j,d,o}(\vec x,t)\,.
\end{equation}

For solving the perturbative ansatz \eqref{eq:second_order_time_dependent_perturbation_theory}, we also need to perform the time integrals at the beginning of the perturbative expression, where we take advantage of the circumstance that the potentials $A_{j,d,o}(\vec x,t)$ with envelope $\eta(t)$ and the phases from the interaction picture $e^{i(E_{a,b}^{\vec \kappa})t/\hbar}$ can all be expanded in terms of time exponentials. This requires a simultaneous sum over each term $A_{j_1,d,o}(\vec x,t_1)$, with polarization index $j_1$, propagation direction $d$ and photon phase oscillation index $o$ of the first interaction $V_\textrm{int}(t_1)$ and each term  $A_{j_2,d',o'}(\vec x,t_2)$ with corresponding indices $j_2,d',o'$, of the second interaction $V_\textrm{int}(t_2)$ in the perturbative calculation.

However, due to the intended scattering geometry, as in Fig. \ref{fig:perturbative_illustration}, with initial electron momentum $-\hbar k_L \vec e_x + k_0 \vec e_y$ and approximate final momentum 
$\hbar k_L \vec e_x + k_0 \vec e_y$, we only account for the Gaussian beam exponentials \eqref{eq:gaussian_beam_exponentials} which are peaked at $\hbar k_L \vec e_x$ in momentum space. This means, that the factor $e^{-iod k_L x}$ from the combined equations \eqref{eq:gaussian_beam_phases} and \eqref{eq:gaussian_beam_exponentials} needs to have a negative product $od$ for the first interaction at $t_1$, as well as $o' d'$ to be negative for the second interaction at $t_2$, which implies the conditions
\begin{equation}
 od=-1\,,\qquad o'd'=-1\,,\label{eq:propagation_direction_constraint}
\end{equation}
for $o,o',d,d' \in \{-1,1\}$. We furthermore only account for resonant combinations of the interaction potentials, in which the time dependent phase $e^{i (o \omega t_1 + o' \omega t_2)}$ is to be constant, as discussed in references \cite{ahrens_bauke_2012_spin-kde,ahrens_bauke_2013_relativistic_KDE,ahrens_2012_phdthesis_KDE,ahrens_2020_two_photon_bragg_scattering,ahrens_guan_2022_beam_focus_longitudinal}. Therefore, we have the additional requirement
\begin{equation}
 o-o'=0 \label{eq:rotating_wave_approximation}
\end{equation}
in our implementation.

We return to the time integration in Eq. \eqref{eq:second_order_time_dependent_perturbation_theory}. Note, that all time-dependent terms are the phases $e^{i (o \omega t_1 + o' \omega t_2)}$ in combination with the envelope $\eta(t)$ in Eq. \eqref{eq:field_temporal_envelope} and the phases $e^{i(E_{a,b}^{\vec \kappa})t/\hbar}$ from the interaction Hamiltonian in the interaction picture \eqref{eq:interaction_matrix_element_computed}. Collecting the time-dependent terms and combining them with the two time-integrations in the perturbative expression \eqref{eq:second_order_time_dependent_perturbation_theory} results in
\begin{multline}\label{eq:double_perturtubation_time_integral}%
\Xi=\int_{0}^{T}d{t_2}\sin^{2}\left({\Omega{t_2}}\right)\exp(i\eta_{A}t_2)\\
\times\int_{0}^{t_2}d{t_1}\sin^{2}\left({\Omega{t_1}}\right)\exp(i\eta_{B}t_1)\,,
\end{multline}%
where we have introduced the abbreviations
\begin{subequations}%
\begin{align}%
\eta_{A}& =\gamma'' E_{a'',b''}^{\vec \kappa''}\hbar^{-1}
- \gamma' E_{a',b'}^{\vec \kappa'}\hbar^{-1} + o' \omega \\
\eta_{B}&=\gamma' E_{a',b'}^{\vec \kappa'}\hbar^{-1}
- \gamma E_{a,b}^{\vec \kappa}\hbar^{-1} + o \omega\,.
\end{align}%
\end{subequations}%
The temporal envelope \eqref{eq:field_temporal_envelope} implies the beginning of the interaction at $t_0=0$ and the ending at $t=T$ in the double time integral \eqref{eq:double_perturtubation_time_integral}. The solution of the integral \eqref{eq:double_perturtubation_time_integral} is discussed in appendix \ref{sec:perturbative_integral_solution}.

\subsection{Iteration scheme of perturbative calculation\label{sec:perturbative_iteration_sceme}}

The perturbative expression in Eq. \eqref{eq:second_order_time_dependent_perturbation_theory} is computed as follows. Based on the description in section \ref{sec:gaussian_beam_potentials}, we obtain the entries for the interaction operator $V_\textrm{int}(t)$ from the vector potential $\vec A'(\vec k^{\vec K},t)$ as illustratively displayed in Fig. \ref{fig:vector_field_momentum_space_magnified}. We then iterate over the following indices
\begin{itemize}
 \item $a_1$: $x$-grid momentum index of first interaction
 \item $b_1$: $y$-grid momentum index of first interaction
 \item $a_2$: $x$-grid momentum index of second interaction
 \item $b_2$: $y$-grid momentum index of second interaction
 \item $s$: initial electron spin
 \item $s'$: intermediate electron spin
 \item $s''$: final electron spin
 \item $\gamma'$: intermediate energy sign of Dirac solutions
 \item $o$: photon absorption/emission of first interaction
 \item $j_1$: field polarization of first interaction
 \item $j_2$: field polarization of second interaction
\end{itemize}
and add the contribution
\begin{multline}\label{eq:perturbative_time_evolution_matrix_elements}
\sum_{a_1,b_1,j_1,\atop{ a_2,b_2,j_2,\atop \gamma',s',o}}\left({u_{\vec{k}_{a'',b''}^{\vec{\kappa}''}}^{\gamma'',s''}}^{\dagger}
\alpha_{j_2} 
u_{\vec{k}_{a',b'}^{\vec{\kappa}'}}^{\gamma',s'} \right)
\left({u_{\vec{k}_{a',b'}^{\vec{\kappa}'}}^{\gamma',s'}}^{\dagger}
\alpha_{j_1} 
u_{\vec{k}_{a,b}^{\vec{\kappa}}}^{\gamma,s}\right)\\
\times A_{j_2}'(k_{a_2,b_2}^{\vec{\mathbf{K}}},0) A_{j_1}'(k_{a_1,b_1}^{\vec{\mathbf{K}}},0) \\
\times \Xi_{a'',a',a;b'',b',b}^{\gamma'',\gamma',\gamma;\sigma'',\sigma',\sigma;o,o'}\,,
\end{multline}
of each combination to the matrix entry \eqref{eq:explicit_propagator_matrix_element} of the time propagation $U(T,0)$. Note, that the index $o'=-o$ for the photon absorption/emission of the second interaction is implied by $o,o' \in \{-1,1\}$ and constraint \eqref{eq:rotating_wave_approximation}. Similarly, the Gaussian beam propagation direction indices $d,d' \in \{-1,1\}$ are implied by the constraints \eqref{eq:propagation_direction_constraint} to be $d=-o$ and $d'=-o'$. We also mention that the solution $\Xi$ of the time integral \eqref{eq:double_perturtubation_time_integral} is a function of the index sets $I, I',I''$, as well as of the parameters $\vec \kappa$, $\vec \kappa'$ and $\vec \kappa''$. We also mention that we set the time parameter in the potentials to $t=0$ and the envelope $\eta(0)=1$, as done in Figs. \ref{fig:vector_potential_momentum_space_high_resolution} and \ref{fig:vector_field_momentum_space_magnified}. The reason for that is that we factor out all of the time-dependence in the potentials, such that the product of all time-dependent functions of in the time-integral \eqref{eq:double_perturtubation_time_integral} can be solved.

\subsection{Spin orientation\label{sec:spin_orientiation}}

The spin dynamics of the Kapitza-Dirac effect in a linear polarized system is characterized by a spin rotation \cite{ahrens_bauke_2013_relativistic_KDE,ahrens_2020_two_photon_bragg_scattering,ahrens_guan_2022_beam_focus_longitudinal}. The references imply that the direction of the spin rotation axis is the normal vector of the plane which is spanned by the laser propagation direction and the laser polarization direction, provided the electron momentum is parallel to this plane as well. Consequently, for the system of consideration here, an electron which is polarized along the $z$-axis will only acquire a phase. Spin effects can only be observed for spin configurations in the $x$-$y$ plane. We therefore require an $x$-polarized spin state for the initial electron quantum state. For the case of the intended momentum eigenstate of positive eigenenergy ($\gamma=+1$) at $a=b=0$ this reads as
\begin{equation}
 c_{0,0}^{+1,0}(0)=\frac{1}{\sqrt{2}}\,,\qquad
 c_{0,0}^{+1,1}(0)=\frac{1}{\sqrt{2}}\,.\label{eq:initial_condition}
\end{equation}
We also project the final quantum state at the at the $x$-quantization axis by introducing the variables
\begin{subequations}\label{eq:final_x_polarization_projection}%
\begin{align}%
 c_{a'',b'';x}^{\gamma'',0}(T) &= \frac{1}{\sqrt{2}}\left[
 c_{a'',b''}^{\gamma'',0}(T)
 + c_{a'',b''}^{\gamma'',1}(T)\right]\\
 c_{a'',b'';x}^{\gamma'',1}(T) &= \frac{1}{\sqrt{2}}\left[
 c_{a'',b''}^{\gamma'',0}(T)
 - c_{a'',b''}^{\gamma'',1}(T)\right]\,.
\end{align}%
\end{subequations}%

\section{Results\label{sec:results}}

\subsection{Simulation setup parameters\label{sec:simulation_parameters}}

\begin{figure}
	\includegraphics[width=0.5 \textwidth]{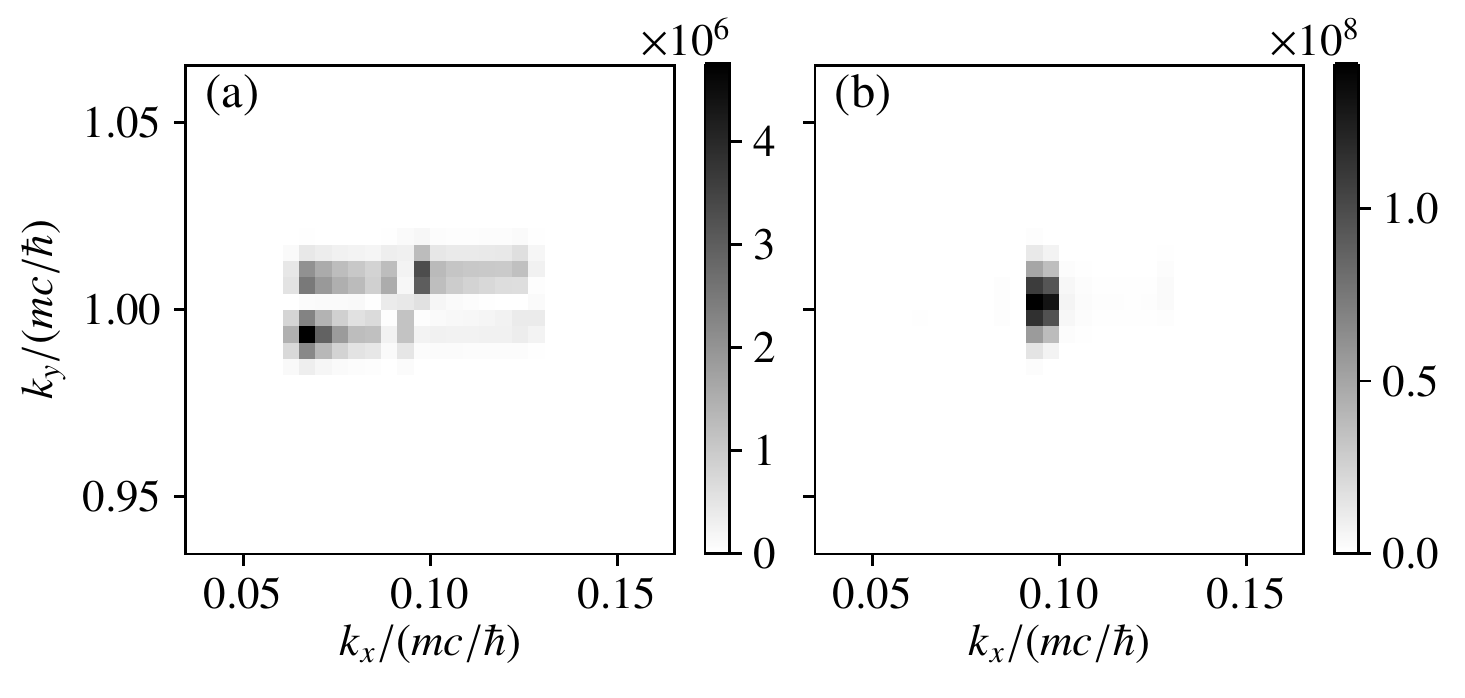}
	\caption{\label{fig:final_diffraction_pattern} Diffraction pattern computed according to our method, as discussed in section \ref{sec:implementation_details}. Displayed are the probabilities $|c_{a'',b'';x}^{+1,0}(T)|^2$ (left panel) and $|c_{a'',b'';x}^{+1,1}(T)|^2$ (right panel) according to Eq. \eqref{eq:final_x_polarization_projection}, with initial condition \eqref{eq:initial_condition}. For our simulation, we set the parameters $T=10^3 \hbar/(mc^2)$ and $k_0=1 mc/\hbar$, with the other parameters implied by sections \ref{sec:gaussian_beam_potentials} and \ref{sec:simulation_parameters}.}
\end{figure}

We are now ready to introduce our simulation result in Fig. \ref{fig:final_diffraction_pattern}. The laser field in this simulation has the properties as described in section \ref{sec:gaussian_beam_potentials} with parameters $k_L=0.1\,mc/\hbar$ for the laser wave number, $\epsilon=0.1$ for the diffraction angle and $T=10^3 \hbar/(mc^2)$ for the duration of the interaction between electron and laser field. For the exploration of our method, we set $A_0=1mc^2/q$ for the field amplitude, which corresponds to a laser intensity of $4.6\times 10^{27}\,\textrm{W}/\textrm{cm}^2$. In this context, we mention that the field amplitude enters the perturbative calculation linearly, which means that $A_0$ in Eq. \eqref{eq:gaussian_beam_exponentials} enters the final expression \eqref{eq:perturbative_time_evolution_matrix_elements} twice, as implied by Eq. \eqref{eq:interaction_operators_application} and could be factored out from all terms. A discussion about suitable parameters for an experimental implementation can be found in Ref. \cite{ahrens_2020_two_photon_bragg_scattering}.

We further set the grid resolution $N_x=N_y=15$ for our computational setup. For the electron, we set the initial momentum $\vec p_\textrm{in}$ as in Eq. \eqref{eq:initial_momentum}, with transverse momentum $k_0=1mc$ and initial $x$-polarized quantum state \eqref{eq:initial_condition}. With the specified properties, Fig. \ref{fig:final_diffraction_pattern} displays the absolute value squared of the final quantum state $|c_{a'',b'';x}^{\gamma'',s}(T)|^2$ as in Eq. \eqref{eq:final_x_polarization_projection}, which is formally implied by the propagation \eqref{eq:general_time_evolution}, for which the contributions \eqref{eq:perturbative_time_evolution_matrix_elements} are summed and iterated according to the scheme as described in section \ref{sec:perturbative_iteration_sceme}.

\subsection{Simulation properties\label{sec:physical_meaning}}

We see in Fig. \ref{fig:final_diffraction_pattern} that the spin-flip amplitude in Fig. \ref{fig:final_diffraction_pattern}(b) is enhanced by about two orders of magnitude, compared to the spin preserving amplitude in \ref{fig:final_diffraction_pattern}(a). This corresponds to the spin-flip in the Kapitza-Dirac effect, as described in references \cite{ahrens_bauke_2013_relativistic_KDE,ahrens_2020_two_photon_bragg_scattering}. It also matches a recent numeric approach of the same setup, which is based on the Fast-Fourier Transform (FFT) split operator method \cite{ge_ahrens_2024_Kapitza_Dirac_simulation}, and also shows a two order of magnitude enhancement of the spin-flip amplitude over the no-flip amplitude in the diffraction peak.

Most interesting is a comparison of the results in Fig. \ref{fig:final_diffraction_pattern} with the perturbative calculation in Ref. \cite{ahrens_guan_2022_beam_focus_longitudinal}, which is the predecessor of the perturbative calculation in this article and provides additional scaling relations for the diffraction effect regarding the laser wavelength and the beam focus. The final diffraction probabilities in Ref. \cite{ahrens_guan_2022_beam_focus_longitudinal} are given in Eq. (80) for a spin-preserving term and Eq. (81) for the spin-flip term, where the latter corresponds to the plane-wave case from earlier literature \cite{ahrens_bauke_2013_relativistic_KDE,ahrens_2020_two_photon_bragg_scattering}. In this context, we need to mention that Eq. (80) in Ref. \cite{ahrens_guan_2022_beam_focus_longitudinal} contains a typo: The variable $\epsilon$ in both square brackets needs to be squared to be correct. Accounting for the $\epsilon^2$ term, we find that the spin-flip in Eq. (81) is enhanced by a factor of 400 over the spin-preserving term in Eq. (80), for the parameters $\epsilon=0.1$ and $q_L=0.1$, which is similar to our finding from Fig. \ref{fig:final_diffraction_pattern}. Here, we need to emphasize that Eq. (80) in Ref. \cite{ahrens_guan_2022_beam_focus_longitudinal} only accounts for the term (77b), which does not alter the electron momentum along the electron propagation direction. There is a spin preserving term in Eq. (76a), which implies a simultaneous momentum change by $q_z=\pm 2 \epsilon q_L$ along the electron propagation direction and another spin preserving term in Eq. (77a), which implies the doubled momentum change $\pm 2 |q_z|$. Both equations, (76a) and (77a), scale dominantly with 'only' $\epsilon^2$, and have the same order of magnitude amplitude as Eq. (77b). Interestingly, the momentum change by $\pm |q_z|$ due to (76a) matches the vertical double peak structure in Fig. \ref{fig:final_diffraction_pattern}(a), with a similar momentum transfer along the electron propagation direction of $\pm q_z = 0.2 m c/\hbar$.

Fig. \ref{fig:final_diffraction_pattern}(a) also contains a horizontal triple peak structure, which corresponds to a momentum transfer of about $\pm 0.25 m c/\hbar$ along the laser propagation direction, additional to the conventional momentum transfer in the Kapitza-Dirac effect. In this context, the attribute 'conventional' refers to the $2 \hbar k_L$ momentum change, from the initial electron momentum $-\hbar k_L$ to the final electron momentum $\hbar k_L$ along the $x$-direction. Note, that a momentum transfer along the laser propagation direction differently from the usual $2 \hbar k_L$ momentum in the Kapitza-Dirac effect has neither been accounted for in the plane-wave investigations \cite{ahrens_bauke_2013_relativistic_KDE,ahrens_2020_two_photon_bragg_scattering}, nor in the perturbative beam focus solution in \cite{ahrens_guan_2022_beam_focus_longitudinal}. However, a merging of the initial electron beam and its diffraction peak is observed for high beam foci in a realistic numerical study of the Kapitza-Dirac effect \cite{ge_ahrens_2024_Kapitza_Dirac_simulation}. One could consider this merging effect as an indication for the mentioned horizontal momentum transfer towards lower momenta along the $k_x$ direction in Fig. \ref{fig:final_diffraction_pattern}(a). Nevertheless, the conclusion about an additional momentum transfer along the laser propagation direction should be taken with care, as the results in this paper are dependent on the grid resolution along the $x$-direction, as described in the section about code convergence in appendix \ref{sec:code_convergence}. Besides discussing the code convergence, we also investigate the influence of the longitudinal laser polarization component on the quantum dynamics in appendix \ref{sec:further_properties}.

\subsection{The negative solutions of the Dirac equation\label{sec:negative_solutions}}

Performing a calculation with a relativistic wave equation for the electron quantum state involves also the negative solutions of the Dirac equation. In the context of Dirac's hole theory \cite{Dirac_1930_hole_theory,wachter_2011_relativistic_quantum_mechanics,schwabl_2000_advanced_quantum_mechanics} one would associate anti-particles with the negative solutions and indeed, the calculation process for computing electron-positron pair-creation involves the propagation from negative to positive energy-eigenstates (or vice versa) \cite{Fradkin_Gitman_Shvartsman_1991_Quantum_Electrodynamics_with_Unstable_Vacuum}.

According to a semi-classical interpretation picture \cite{ruf_2009_pair_creation,ahrens_bauke_2012_spin-kde,ahrens_bauke_2013_relativistic_KDE}, the photon energy $c \hbar k_L = 0.1 m c^2$ of the parameters in the present investigation would require the interaction with at least 20 laser photons, for bridging the mass-gap from the negative solutions to the positive solutions. Resonances with 4 interacting photons \cite{Hu_2020_multi_channel_pair_creation} up to 11 interacting photons \cite{ruf_2009_pair_creation,ruf_2010_pair_creation_long} are reported. Note, that one needs one perturbative order for each interacting photon. Consequently, the second order perturbative solution in this paper would be capable for describing the laser pair-creation effect of the Breit-Wheeler process \cite{Breit_Wheeler_1934_pair_creation}, in which two photons are converted into an electron-position pair. We therefore assume that the process of pair-creation contributes only a vanishingly small effect to the process of electron diffraction which discussed here. This conclusion is underlined by the values $\xi=1$ for the classical nonlinearity parameter and $\chi=0.1$ for quantum nonlinear parameter \cite{Di_Piazza_2012_strong_field_review,fedotov_2023_strong_field_review}, which one would identify for the setup in this paper.

This doesn't mean that the presence of negative energy states is irrelevant for our topic. Evidence can be seen in the deduction of the non-relativistic limit of the Dirac equation, resulting in the Pauli equation \cite{wachter_2011_relativistic_quantum_mechanics,schwabl_2000_advanced_quantum_mechanics}, which involves also the lower components of the Dirac equation. On the other hand, numeric solutions of the Dirac equation for sub-critical external fields \cite{ahrens_bauke_2012_spin-kde,ahrens_bauke_2013_relativistic_KDE,bauke_ahrens_2014_spin_precession_1,bauke_ahrens_2014_spin_precession_2,ahrens_2017_spin_filter,ahrens_2020_two_photon_bragg_scattering,ge_ahrens_2024_Kapitza_Dirac_simulation} generally display low amplitudes of negative states after field turn off. In the context of these considerations we mention that our calculation is naturally involving the negative solutions of the Dirac equation. In particular the intermediate index set $I'$ in Eq. \eqref{eq:interaction_operators_application} also includes the particle energy sign $\gamma'$ which persists until the finally evaluated expression in Eq. \eqref{eq:perturbative_time_evolution_matrix_elements}. However, a quantitative analysis of the contribution of the negative states is appearing complicated due to the summation over all possible quantum paths in Eq. \eqref{eq:perturbative_time_evolution_matrix_elements}. An incoherent sum over the absolute value squares of each of the terms in \eqref{eq:perturbative_time_evolution_matrix_elements} yields a significant dominance for the sum of terms with $\gamma'=1$, compared to the sum of terms with $\gamma'=-1$.

\subsection{Code execution time scaling\label{sec:execution_scaling}}

In this section we want to quantify the performance of our method and its scaling with respect to numerical and physical parameters. For that we recall that the algorithm is based on the perturbative expression \eqref{eq:second_order_time_dependent_perturbation_theory} with its operator application expansion \eqref{eq:interaction_operators_application}, which finally results in the explicit perturbative contributions \eqref{eq:perturbative_time_evolution_matrix_elements}, which contain iteration indices as listed in section \ref{sec:perturbative_iteration_sceme}. An extension of this scheme to $n_D$ spacial dimensions will require and iteration over the $n_D$ grid indices of each dimension for each perturbative order. For example in three dimensions with $n_D=3$ the number of grid indices would be $N_x=N_1$, $N_y=N_2$, $N_z=N_3$ for the grid points along the $x$, $y$, $z$ dimension. In $n_D$ dimensions, the vector potential also has $n_D$ vector components, over which one has to iterate over. And for each perturbative order one also needs to iterate over the electron polarization degree of freedom, consisting of spin (factor of 2) and sign of energy (factor of 2). Note, that also the initial state of the electron has these $2\times2$ degrees of freedom. Further, one may either choose to setup the initial electron quantum state as a momentum eigenstate as it is done in this article or to initialize the wave function on an $n_D$ dimensional grid, for example as Gaussian wave packet. Taking all factors together results in the total number of elements to be iterated over
\begin{equation}
 4 \left(4 n_D \right)^{n_p}\left(\prod_{n_d=1}^{n_D} N_{n_d}\right)^{n_p + n_I}\,,\label{eq:number_of_summation_elements}
\end{equation}
over which one needs to iterate over, for computing the electron diffraction pattern. The number $n_p$ Eq. \eqref{eq:number_of_summation_elements}  is the number of interacting photons in the Kapitza-Dirac interaction, which corresponds to the number of orders in perturbation theory, for computing the herein considered general type of $n_p$-photon Kapitza-Dirac effect. The number $n_I$ represents the initial condition, where $n_I=0$ corresponds to an initial momentum eigenstate and we set $n_I=1$ for the setup of an initial electron quantum state on the $n_D$ dimensional grid. Note, that Eq. \eqref{eq:number_of_summation_elements} is not including the general combinatorial possibilities of photon absorptions and photon emissions, corresponding to the index $o$ in the text above.

For the specific scenario in this work, we have a two-photon interaction $n_p=2$ in a two-dimensional setup $n_D=2$ with a momentum eigenstate as initial condition $n_I=0$. Accounting for an initial and final positive energy state (omission of two factors of 2) and the possible photon absorption and emission combinations (factor of two for the possible configurations of the variable $o$), one obtains from Eq. \eqref{eq:number_of_summation_elements} the number of iterations $2^7 N_x^2 N_y^2=6.48\cdot10^6$, for the given parameters $N_x=N_y=15$. On our machine, we obtain a runtime of $284.3$ seconds, implying an average computation time of $43.9\,\mu\textrm{s}$ for each of the contributions \eqref{eq:perturbative_time_evolution_matrix_elements}.

\section{Discussion and outlook\label{sec:discussion_and_outlook}}

In this article we have been investigating a time-propagation method for evolving quantum states in time on the basis of time-dependent perturbation theory, for simulating the spin resolved two-photon Kapitza-Dirac effect with a realistic Gaussian beam shaped laser pulse in two dimensions. The method is solving the time-evolution by the integration of phases which arise from different quantum state energies at initial, intermediate and final times in momentum space. The potential of the external laser field in momentum space is thereby obtained from a Fourier transform and is thus exceeding the spacial accuracy of previous computations of Kapitza-Dirac scattering. Compared to other numerical methods, for example the FFT-split operator method \cite{bauke_2011_GPU_acceleration_FFT_split_operator,ge_ahrens_2024_Kapitza_Dirac_simulation}, our procedure has the advantage to not iterate over a large number of time steps, because the time-dependence is computed analytically (see appendix \ref{sec:perturbative_integral_solution}). Specifically, our method is computing the quantum time-evolution without time step iteration and appears beneficial in particular for our focus on relativistic quantum solutions. The reason is that relativistic descriptions unavoidably contain a mass gap in the energy-momentum relation, which in turn implies the necessity of small and thus numerous time steps.

A further advantage of the perturbative solution technique is an iteration over all possible quantum state configurations, which allows for arbitrary parallelization of the numeric implementation with low effort. However, disadvantageous is the unbeneficial scaling of the computation time with grid resolution in space, as discussed in the previous section \ref{sec:execution_scaling}. The runtime scales as $O(N_x^2 N_y^2)$ already for most simple scenario of a two-photon interaction with an initial momentum eigenstate, which is the use case which we have presented here in this article. Such a scaling is worse than the scaling of the split-operator method, which scales as $O[N_x \log(N_x) N_y \log(N_y)]$.

On the other hand, our perturbative solution technique is designed such that the position- and momentum space grid can be adjusted to the specific interaction regions, as described in section \ref{sec:gaussian_beam_potentials} and illustrated in Figs. \ref{fig:vector_potential_momentum_space_high_resolution}, \ref{fig:vector_field_momentum_space_magnified} and \ref{fig:perturbative_illustration}, which might partially compensate a disadvantageous runtime scaling. The possibility of running the implementation at a low grid resolution is impressively demonstrated by the convergence of the simulation result along the $y$-dimension for only $N_y=15$ grid points (see appendix \ref{sec:code_convergence} for comments on code convergence). At the same time, the non-convergent behavior along the $x$-dimension for the tested resolutions is the biggest drawback of our method and calls its use into question. We attribute the lack of simulation convergence along the $x$-axis to the non-trivial phase \eqref{eq:gaussian_beam_phases} of the Gaussian beam. We may propose a solution approach for fixing the lack of convergence by having a look at the construction procedure of the Gaussian beam in momentum space \cite{Quesnel_1998_gaussian_beam_coulomb_gauge}. We see in reference \cite{Quesnel_1998_gaussian_beam_coulomb_gauge} that the Gaussian beam field in momentum space is only non-vanishing at the photon dispersion
\begin{equation}
\omega = c |\vec k|\,,\label{eq:photon_dispersion}
\end{equation}
in agreement to what one can observe in Figs. \ref{fig:vector_potential_momentum_space_high_resolution} and \ref{fig:vector_field_momentum_space_magnified}.
Therefore, it appears tempting to parameterize the potential of the Gaussian beam only along the constraint \eqref{eq:photon_dispersion} in momentum space. Unfortunately, such a procedure is not compatible with a Fourier transform on a rectangular grid space, such that the solution method as described in this article would have to be modified significantly. One advantage of parameterizing the potential along the dispersion \eqref{eq:photon_dispersion} would be the additional constraint \eqref{eq:photon_dispersion} on the $n_D$ dimensional simulation space such that only a $n_D-1$ dimensional hypersurface needs to be parameterized. For example, the two-photon scattering of a momentum eigenstate in a two-dimensional system, corresponding to the scenario which is discussed in this article, only requires a one-dimensional grid with $N$ grid points, implying a corresponding reduced runtime scaling $O(N^2)$.

\begin{acknowledgments}
The work was supported by the National Natural Science Foundation of China (Grants No. 11975155 and 11935008).
\end{acknowledgments}

\appendix

\section{Solution of the perturbative double time integration\label{sec:perturbative_integral_solution}}

For solving the double time integral in Eq. \eqref{eq:double_perturtubation_time_integral} we first introduce the variables
\begin{subequations}%
\begin{align}%
\eta_{A\pm1}& =\eta_{A}\pm 2 \Omega \\
\eta_{B\pm1}& =\eta_{B}\pm 2 \Omega\,,
\end{align}%
\end{subequations}%
such that we can expand the integral into
\begin{multline}\label{eq:perturtubation_integral_expanded}%
\Xi=\int_{0}^{T}d{t_2}\Bigg[ \frac{1}{2}\exp(i\eta_{A}t_2)
-\frac{1}{4}\ \exp(i\eta_{A+1}t_2)\\
-\frac{1}{4} \exp(i\eta_{A-1}t_2) \Bigg] \int_{0}^{t_2}d{t_1} \Bigg[ \frac{1}{2}\exp(i\eta_{B}t_1)\\
-\frac{1}{4}\ \exp(i\eta_{B+1}t_1)
-\frac{1}{4} \exp(i\eta_{B-1}t_1) \Bigg]\,.
\end{multline}%
For the integration over $t_1$ we introduce further variables
\begin{subequations}%
\begin{align}%
 \zeta_{\phantom{\pm1}} &= \eta_{A}+ \eta_{B}=\gamma''E_{a'',b''}^{\vec \kappa''}- \gamma E_{a,b}^{\vec \kappa}+(o+o') \omega\\
\zeta_{\pm1}& =\zeta \pm 2 \Omega \\
\zeta_{\pm2}& =\zeta \pm 4 \Omega\,.
\end{align}%
\end{subequations}%
After the integration of \eqref{eq:perturtubation_integral_expanded} over $t_1$ and expanding all terms in the square brackets we obtain
\begin{widetext}
\begin{subequations}%
\begin{align}%
\Xi=&\int_{0}^{T}d{t_2}\sin^{2}\left({\Omega{t_2}}\right)\exp(i\eta_{A}t_2)
\int_{0}^{t_2}d{t_1}\sin^{2}\left({\Omega{t_1}}\right)\exp(i\eta_{B}t_1)\\
&\int_{0}^{T}d{t_2} \left[ 
 \frac{\exp(i\zeta t_2)-\exp(i\eta_{A}t_2)}{4i\eta_B}
-\frac{\exp(i\zeta_{+1}t_2)-\exp(i\eta_{A}t_2)}{8i\eta_{B+1}}
-\frac{\exp(i\zeta_{-1}t_2)-\exp(i\eta_{A}t_2)}{8i\eta_{B-1}} \right] \\
-&\int_{0}^{T}d{t_2} \left [
 \frac{\exp(i\zeta_{+1}t_2)-\exp(i\eta_{A+1}t_2)}{8i\eta_{B}}
-\frac{\exp(i\zeta_{+2}t_2)-\exp(i\eta_{A+1}t_2)}{16i\eta_{B+1}}
-\frac{\exp(i\zeta t_2)-\exp(i\eta_{A+1}t_2)}{16i\eta_{B-1}}  \right ] \\
-&\int_{0}^{T}d{t_2} \left[ 
 \frac{\exp(i\zeta_{-1}t_2)-\exp(i\eta_{A-1}t_2)}{8i\eta_{B}}
-\frac{\exp(i\zeta t_2)-\exp(i\eta_{A-1}t_2)}{16i\eta_{B+1}}
-\frac{\exp(i\zeta_{-2} t_2)-\exp(i\eta_{A-1}t_2)}{16i\eta_{B-1}} \right]\,.
\end{align}%
\end{subequations}%
Factorizing all terms with different exponentials yields
\begin{subequations}%
\begin{align}%
\Xi=&\int_{0}^{T}d{t_2}\left[ \exp(i\zeta t_2) \left(\frac{1}{4i\eta_{B}}+ \frac{1}{16i\eta_{B-1}}+\frac{1}{16i\eta_{B+1}} \right )  \right] 
-\int_{0}^{T}d{t_2}\left[ \exp(i\zeta_{+1} t_2) \left( \frac{1}{8i\eta_{B}}+\frac{1}{8i\eta_{B+1}} \right)  \right]\\
-&\int_{0}^{T}d{t_2}\left[ \exp(i\zeta_{-1} t_2) \left(\frac{1}{8i\eta_{B}}+\frac{1}{8i\eta_{B-1}} \right)  \right]
+\int_{0}^{T}d{t_2}\left[ \exp(i\zeta_{+2} t_2) \left(\frac{1}{16i\eta_{B+1}} \right)  \right]\\
+&\int_{0}^{T}d{t_2}\left[ \exp(i\zeta_{-2} t_2) \left(\frac{1}{16i\eta_{B-1}} \right)  \right]
-\int_{0}^{T}d{t_2}\left[ \exp(i\eta_{A} t_2) \left( \frac{1}{4i\eta_{B}}-\frac{1}{8i\eta_{B-1}}-\frac{1}{8i\eta_{B+1}} \right)  \right]\\
+&\int_{0}^{T}d{t_2}\left[ \exp(i\eta_{A+1} t_2) \left( \frac{1}{8i\eta_{B}}-\frac{1}{16i\eta_{B-1}}-\frac{1}{16i\eta_{B+1}} \right)  \right]\\
+&\int_{0}^{T}d{t_2}\left[ \exp(i\eta_{A-1} t_2) \left( \frac{1}{8i\eta_{B}}-\frac{1}{16i\eta_{B-1}}-\frac{1}{16i\eta_{B+1}} \right)  \right]\,.
\end{align}%
\end{subequations}%
Performing the second integral over $t_2$ results in
\begin{subequations}%
\begin{align}%
\Xi=&\frac{1}{i\zeta}\left ( \frac{1}{4i\eta_{B}}+\frac{1}{16i\eta_{B-1}}+\frac{1}{16i\eta_{B+1}}  \right )\left[ \exp(i\zeta T)-1 \right]
- \frac{1}{i\zeta_{+1}} \left ( \frac{1}{8i\eta_{B}}+\frac{1}{8i\eta_{B+1}} \right )\left [ \exp(i\zeta_{+1} T)-1   \right ]\\
-&\frac{1}{i\zeta_{-1}} \left ( \frac{1}{8i\eta_{B}}+\frac{1}{8i\eta_{B-1}} \right )\left [ \exp(i\zeta_{-1} T)-1   \right ]
+\frac{1}{i\zeta_{+2}} \left (\frac{1}{16i\eta_{B+1}} \right )\left [ \exp(i\zeta_{+2} T)-1   \right ]\\
+&\frac{1}{i\zeta_{-2}} \left (\frac{1}{16i\eta_{B-1}} \right )\left [ \exp(i\zeta_{-2} T)-1   \right ]
-\frac{1}{i\eta_{A}}\left ( \frac{1}{4i\eta_{B}}-\frac{1}{8i\eta_{B-1}}-\frac{1}{8i\eta_{B+1}} \right )\left [ \exp(i\eta_{A} T)-1   \right ]\\
+&\frac{1}{i\eta_{A+1}}\left ( \frac{1}{8i\eta_{B}}-\frac{1}{16i\eta_{B-1}}-\frac{1}{16i\eta_{B+1}} \right ) \left [ \exp(i\eta_{A+1} T)-1   \right ]\\
+&\frac{1}{i\eta_{A-1}} \left ( \frac{1}{8i\eta_{B}}-\frac{1}{16i\eta_{B-1}}-\frac{1}{16i\eta_{B+1}} \right ) \left [ \exp(i\eta_{A-1} T)-1   \right ]\,.
\end{align}%
\end{subequations}%
\end{widetext}
Note, that the variables $\zeta$, $\zeta_\pm1$, $\zeta_\pm2$, $\eta_A$ and $\eta_{A\pm1}$ may obtain values close or equal to zero, as we describe the Kapitza-Dirac effect on resonance, as discussed in references \cite{ahrens_bauke_2012_spin-kde,ahrens_bauke_2013_relativistic_KDE,ahrens_guan_2022_beam_focus_longitudinal}. We therefore approximate the integrals in the following way
\begin{equation}%
\int_{0}^{T}d{t_2} \exp(i \zeta t_2)
=\sum_{n=0}^{\infty}\frac{i^{n}}{(n+1)!} \zeta^{n}T^{n+1}
\approx T+ \frac{i \zeta}{2!}T^{2}\,,
\end{equation}%
as shown here for the example of $\zeta$. We choose this form for $\zeta<4.6\cdot10^{-10}$, because higher powers of $\zeta$ are giving only negligibly small contributions.

\section{Code convergence\label{sec:code_convergence}}

For explaining the situation, we first present examples for a situation in which the simulation results are converged, when changing the grid resolution along the $y$-direction, which we show in Figs. \ref{fig:position_y_grid_variation} and \ref{fig:momentum_y_grid_variation}. After Figs. \ref{fig:position_y_grid_variation} and \ref{fig:momentum_y_grid_variation}, we discuss the change of simulation results when changing the grid resolution along the $x$-direction in Figs. \ref{fig:x_grid_variation_up}-\ref{fig:x_grid_variation_down_zero_longitudinal}.

In Fig. \ref{fig:position_y_grid_variation}, we increase the grid resolution along the $y$-dimension in the form $N_y=2^g-1$, with $g\in\{4,5,6,7\}$ from the left to the right. This implies that the grid resolution doubles approximately when moving from the left most panel to the panels in the right. We subtract the numbers $N_y$ by $-1$ in the expression $2^g-1$ for creating the resolution odd numbered. Note, that Eq. \eqref{eq:momentum_space_width} implies a change of the momentum space width $k_{y,w}$, when changing the grid resolution $N_y$, such that the momentum space limits along the $y$-axis change as well. However, in Fig. \ref{fig:position_y_grid_variation} we show the diffraction pattern in the same subregion as in Fig. \ref{fig:final_diffraction_pattern}, such that the simulation results of different grid resolutions can be compared easily. We observe that the diffraction pattern appears the same, when changing the resolution $N_y$ of the $y$-axis and therefore conclude that the simulation is converged with respect to grid resolution changes along the $y$-axis in position space.

It is also possible to vary the grid resolution in momentum space instead of position space. This can be done by combining the change of the grid resolution $N_y$ with a simultaneous change of the simulation box width $y_w$. Specifically, we set the grid resolution $N_y=2^g-1$ with $g\in\{4,5,6,7\}$ and simultaneously set the simulation box width $y_w=10\,\lambda\,2^{g-3}$. For this scheme, Eq. \eqref{eq:momentum_space_width} implies an approximately constant simulation box width in momentum space. The resulting variation of diffraction patterns is displayed in Fig. \ref{fig:momentum_y_grid_variation}, and we observe that the simulation results appear more and more smooth along the $y$-direction. However, the diffraction pattern remains qualitatively unchanged and we therefore conclude convergence of our method along the $y$-axis in momentum space.

While Figs. \ref{fig:position_y_grid_variation} and \ref{fig:momentum_y_grid_variation} are showing examples for code convergence for the grid along the $y$-direction, Figs. \ref{fig:x_grid_variation_up} and \ref{fig:x_grid_variation_down} display a different situation, when varying the grid along the $x$-direction. For studying the situation comprehensively, we vary number of grid points $N_x$ and the width in position space $x_w$ independently from each other, resulting in the triangular panel arrangement in the $4 \times 4$ basis structure in Figs. \ref{fig:x_grid_variation_up} and \ref{fig:x_grid_variation_down}. Due to this panel arrangement we split the results into spin-up results in Fig. \ref{fig:x_grid_variation_up} and spin-down results in Fig. \ref{fig:x_grid_variation_down}. The change of grid resolution for Figs. \ref{fig:x_grid_variation_up} and \ref{fig:x_grid_variation_down} is implemented in the following layout:

We start with a grid resolution as in Fig. \ref{fig:final_diffraction_pattern} in the upper right panel of Figs. \ref{fig:x_grid_variation_up} and \ref{fig:x_grid_variation_down}. Moving from the top panels down to the bottom panels, we double the grid resolution $N_x$ according to the scheme $N_x=2^g-1$ while simultaneously also doubling the simulation box width in position space $x_w=10\,\lambda\,2^{g-3}$. In this way, the $x$-axis in momentum space remains approximately unchanged along the panels aligned in columns in Figs. \ref{fig:x_grid_variation_up} and \ref{fig:x_grid_variation_down}. Thus, the $x$ resolution change along the column panels in Figs. \ref{fig:x_grid_variation_up} and \ref{fig:x_grid_variation_down} corresponds to the $y$ resolution change in Fig. \ref{fig:momentum_y_grid_variation}. When moving from the right panels to the left panels in Figs. \ref{fig:x_grid_variation_up} and \ref{fig:x_grid_variation_down}, we halve the simulation box width in position space $x_w$, such that the diagonally aligned panels share the identical position space width $x_w$.

When looking at the resulting simulations in Figs. \ref{fig:x_grid_variation_up} and \ref{fig:x_grid_variation_down}, we see that the diffraction patterns vary drastically for different resolutions along the $x$-direction in position and momentum space. This is different to the converged behavior along the $y$-axis in Figs. \ref{fig:position_y_grid_variation} and \ref{fig:momentum_y_grid_variation}.

\section{Influence of the longitudinal polarization component\label{sec:further_properties}}

Beam focusing implies a longitudinal polarization component \eqref{eq:gaussian_beam_exponentials_longitudinal} in the Gaussian beam, which was speculated to have influence on the quantum dynamics of the Kapitza-Dirac effect \cite{ahrens_guan_2022_beam_focus_longitudinal}. We are interested in investigating the influence of the longitudinal polarization component and repeat the calculation in Figs. \ref{fig:x_grid_variation_up} and \ref{fig:x_grid_variation_down} again with $A_x$ in Eq. \eqref{eq:gaussian_beam_exponentials_longitudinal} set to zero, resulting in Figs. \ref{fig:x_grid_variation_up_zero_longitudinal} and \ref{fig:x_grid_variation_down_zero_longitudinal}. We observe, that one cannot see a difference between the simulations with and without the longitudinal component. This observation is also confirmed by looking at the maximum values in each of the figure panels, which are listed in table \ref{tab:max_values}. One can see in table \ref{tab:max_values}, that the discrepancy between the maximum values with and without longitudinal component can be found only in the digits after the period or less. Therefore, we conclude that the longitudinal polarization component of Gaussian beams has only marginal influence on the quantum dynamics of the two-photon Kapitza-Dirac effect with parameters $\epsilon=0.1$ and $k_L = 0.1\,m c/\hbar$. This is consistent with conclusions which one can draw from the results in \cite{ahrens_guan_2022_beam_focus_longitudinal} and also the recent numerical study \cite{ge_ahrens_2024_Kapitza_Dirac_simulation}.

\begin{widetext}
\begin{turnpage}
\begin{figure}[H]
\includegraphics[width=1.0 \textheight]{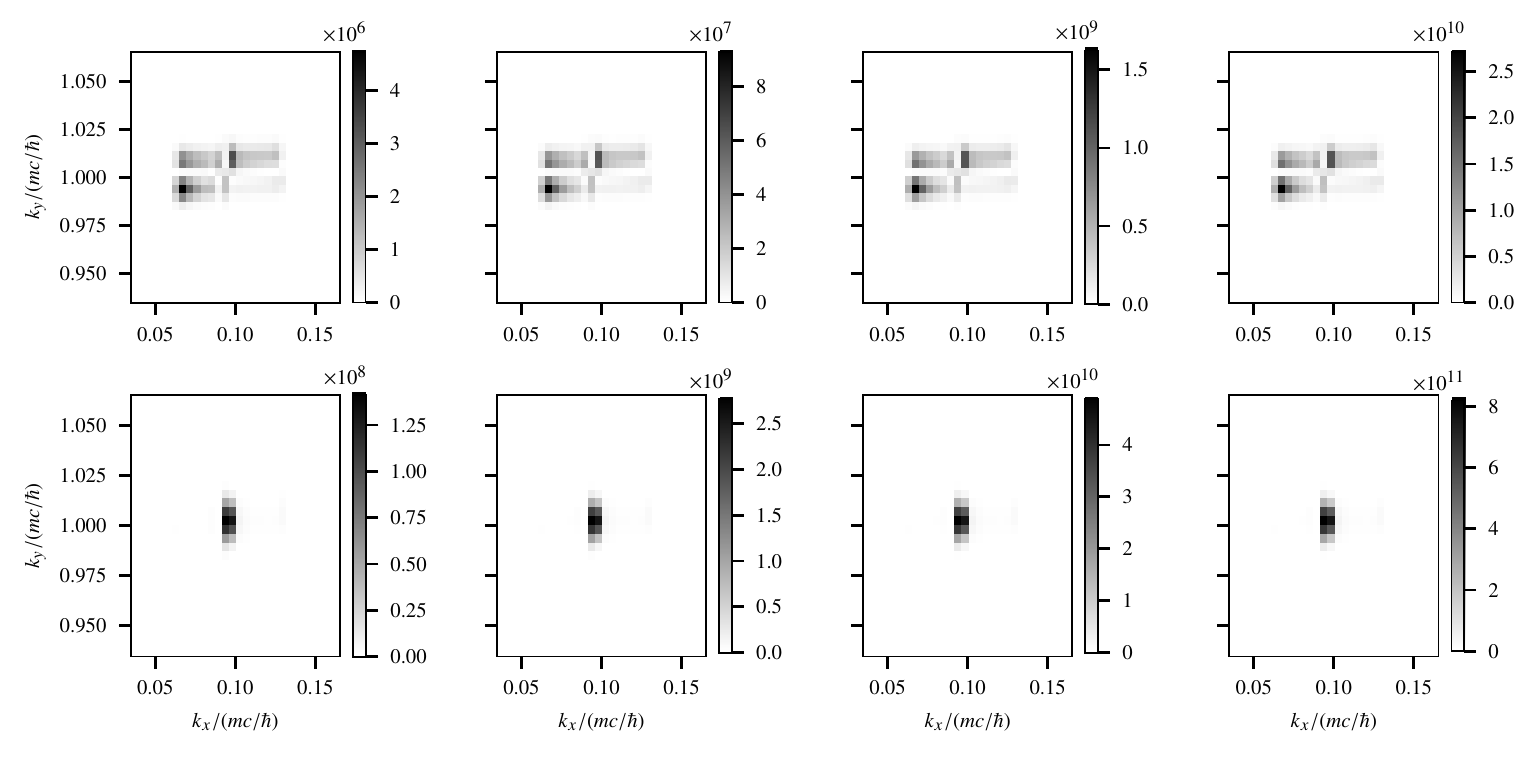}% Here is how to import EPS art
\caption{\label{fig:position_y_grid_variation}Variation of position space grid resolution along the $y$-axis. The upper left panel (spin-up) coincides with Fig. \ref{fig:final_diffraction_pattern}(a) and the lower left panel (spin-down) coincides with Fig. \ref{fig:final_diffraction_pattern}(b). From the left panels to the right panels we increase the grid resolution in position space with the values $N_y\in\{15,31,63,127\}$. The momentum space limits along the $y$-dimension increase accordingly, due to Eq. \eqref{eq:momentum_space_width}. In order to compare the diffraction pattern easier, we display the same momentum space sub-grid-region as in the left-most figures. One can observe that the patterns are not changing their shape, when changing the $y$ grid resolution in position space.}
\end{figure}
\end{turnpage}
\end{widetext}

\begin{widetext}
\begin{turnpage}
\begin{figure}[H]
\includegraphics[width=1.0 \textheight]{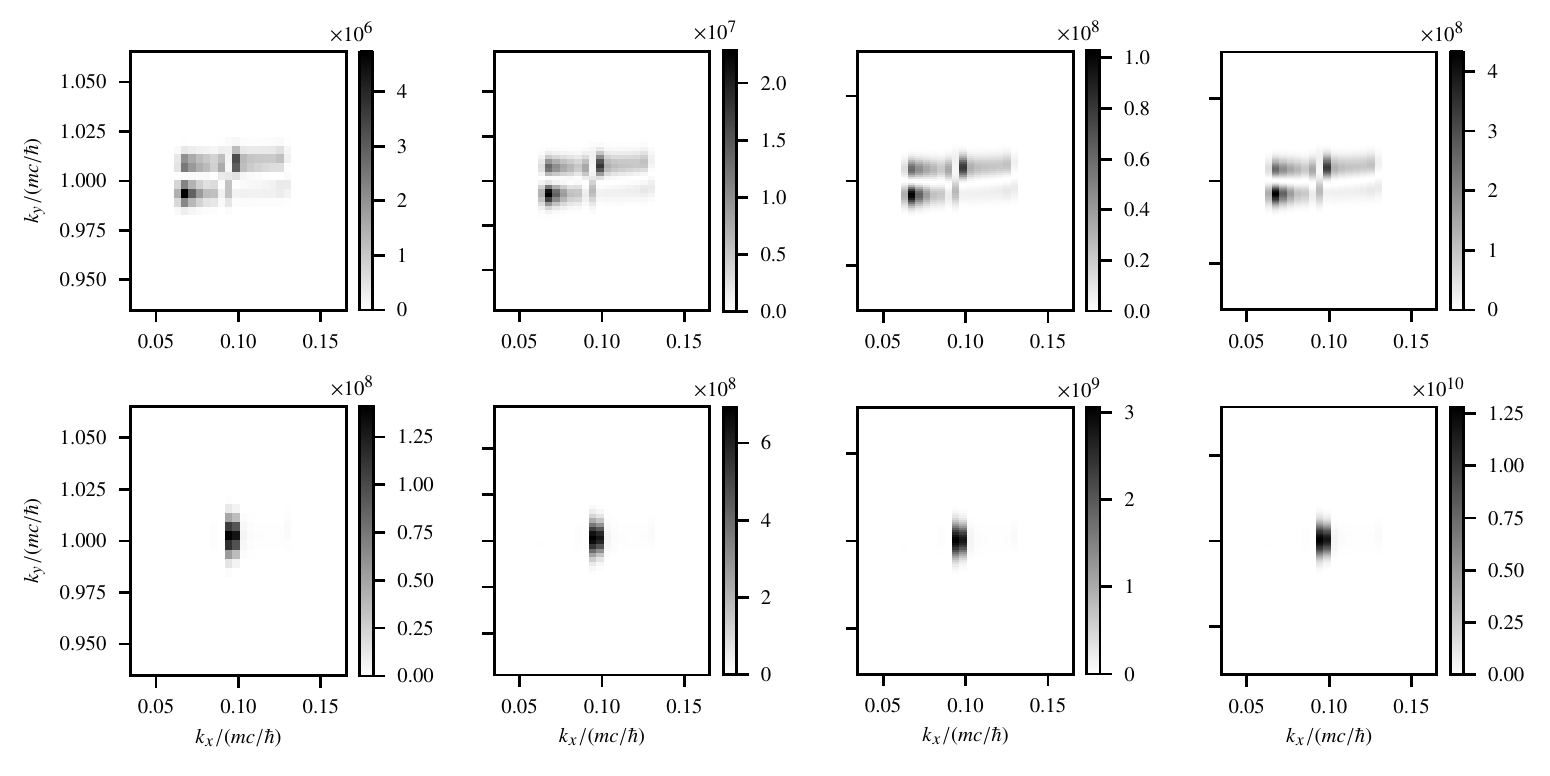}% Here is how to import EPS art
\caption{\label{fig:momentum_y_grid_variation}Variation of momentum space grid resolution along the $y$-axis. As in Fig. \ref{fig:position_y_grid_variation}, the upper left panel (spin-up) coincides with Fig. \ref{fig:final_diffraction_pattern}(a) and the lower left panel (spin-down) coincides with Fig. \ref{fig:final_diffraction_pattern}(b). In contrast to Fig. \ref{fig:position_y_grid_variation} we simultaneously change the grid resolution $N_y\in\{15,31,63,127\}$ and the simulation box width $y_w\in\{20\lambda,40\lambda,80\lambda,160\lambda\}$, when moving from the left to the right panels. According to Eq. \eqref{eq:momentum_space_width}, the limits in momentum space remain approximately unchanged, such that the change of grid points $N_y$ implies a resolution change in momentum space. The diffraction patterns appear similar from the left to the right, besides a smoothing effect along the $y$-direction due to the resolution change.}
\end{figure}
\end{turnpage}
\end{widetext}

\begin{widetext}
\begin{turnpage}
\begin{figure}[H]
\includegraphics[height=0.7 \textwidth]{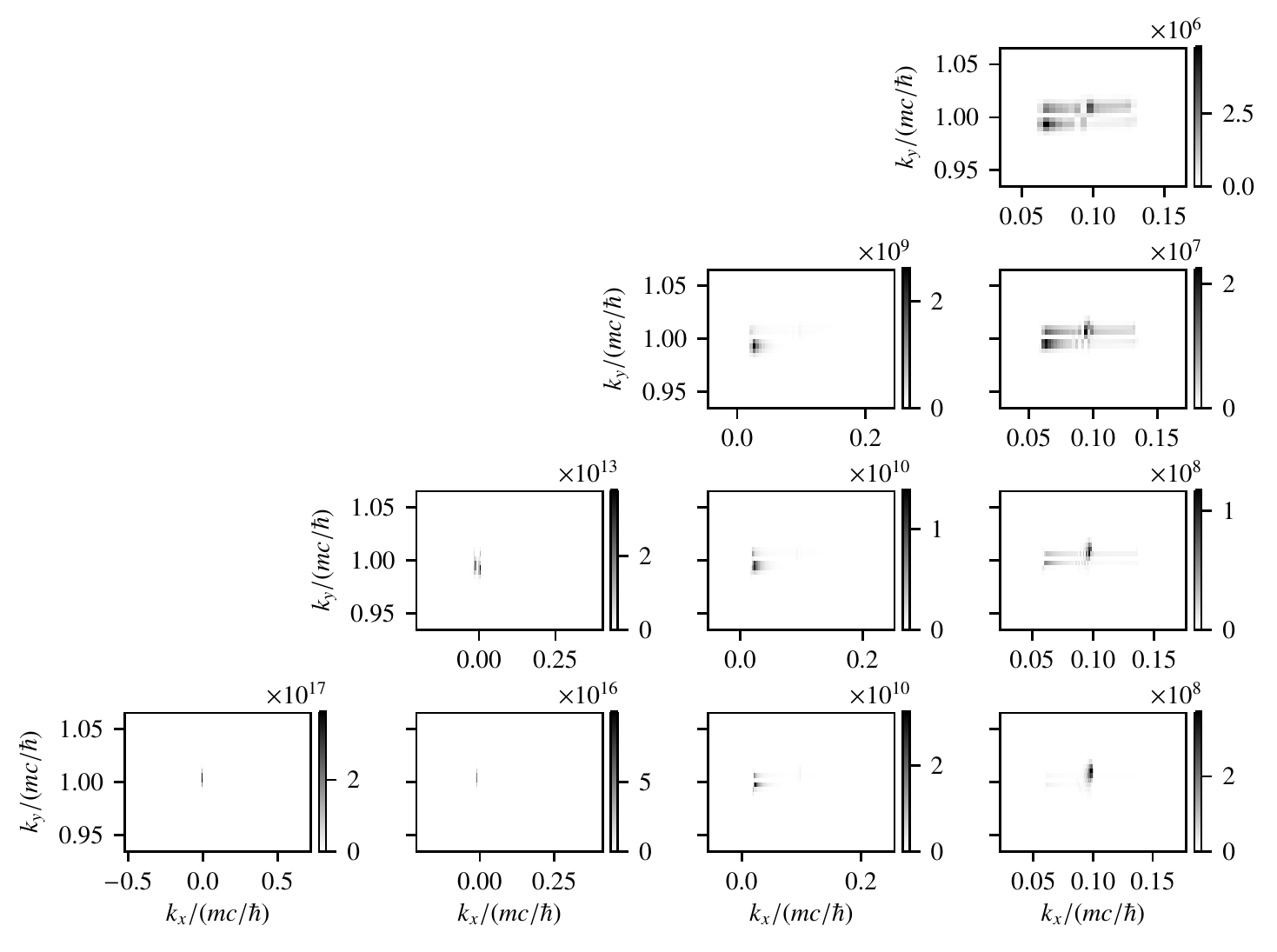}% Here is how to import EPS art
\caption{\label{fig:x_grid_variation_up}Spin-up variation of grid resolution along the $x$-axis. The upper right panel coincides with the spin-up diffraction pattern of Fig. \ref{fig:final_diffraction_pattern}(a). From the top to the bottom panels of the right-most column, we are doubling the resolution according to $N_y\in\{15,31,63,127\}$ and simultaneously we also double simulation box width as $x_w\in\{20\lambda,40\lambda,80\lambda,160\lambda\}$. As a consequence, the limits in momentum space along the $x$-axis remain approximately constant for the right-most panel column, according to Eq. \eqref{eq:momentum_space_width}. From right to the left panels we then halve the simulation box width along the $x$-axis. Therefore, the momentum space resolution is increasing along the panel columns and the position space resolution is increasing along the diagonal panel chains. We observe that the diffraction pattern is changing non-uniformly on grid resolution change.}
\end{figure}
\end{turnpage}
\end{widetext}

\begin{widetext}
\begin{turnpage}
\begin{figure}[H]
\includegraphics[height=0.7 \textwidth]{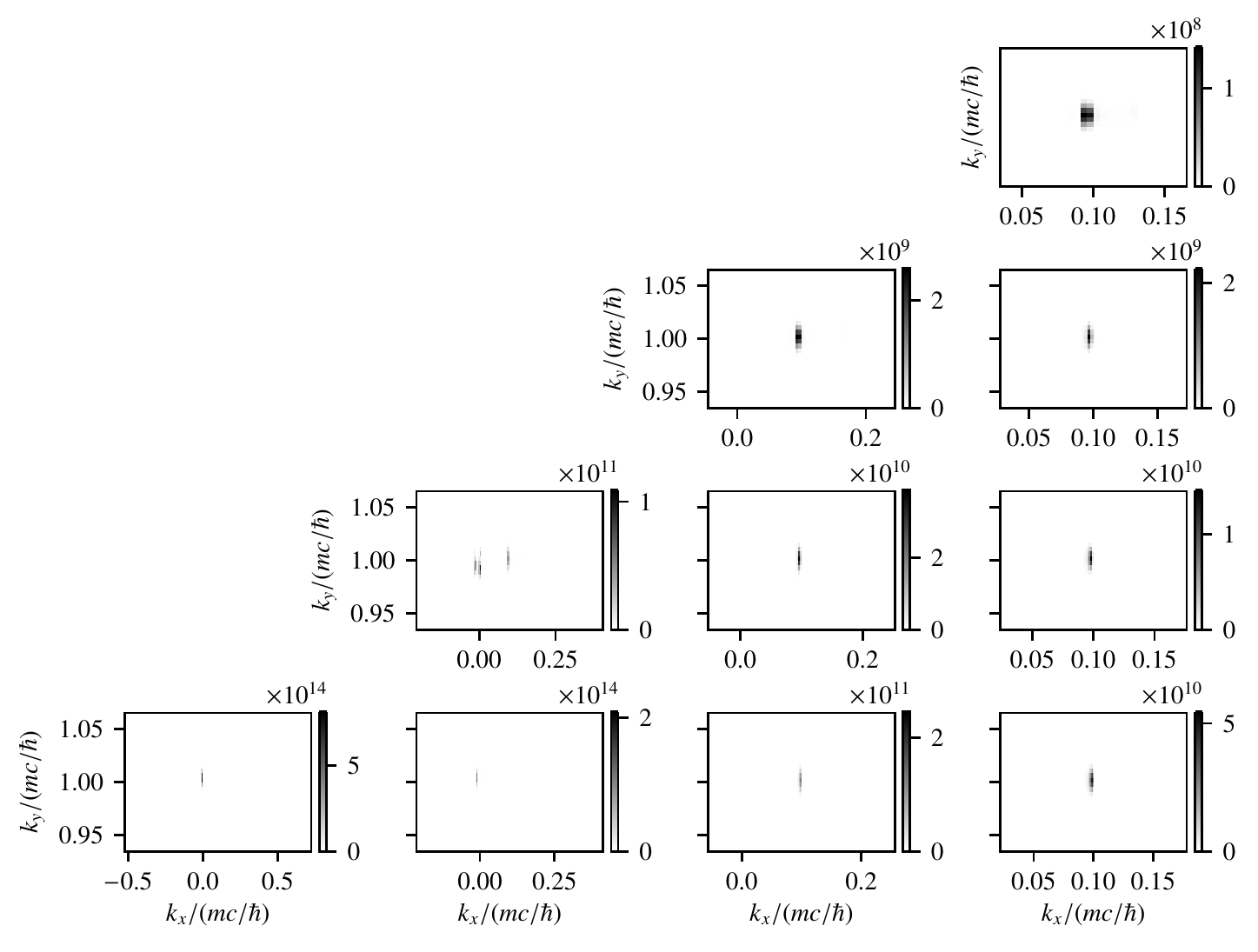}% Here is how to import EPS art
\caption{\label{fig:x_grid_variation_down}Spin-down variation of grid resolution along the $x$-axis. The upper right panel coincides with the spin-down diffraction pattern of Fig. \ref{fig:final_diffraction_pattern}(b). The simulation resolution parameters are changed as in Fig. \ref{fig:x_grid_variation_up}. Also for the spin-down situation we observe a non-uniform change of the diffraction patterns when changing the grid resolution along the $x$-axis.}
\end{figure}
\end{turnpage}
\end{widetext}

\begin{widetext}
\begin{turnpage}
\begin{figure}[H]
\includegraphics[height=0.7 \textwidth]
{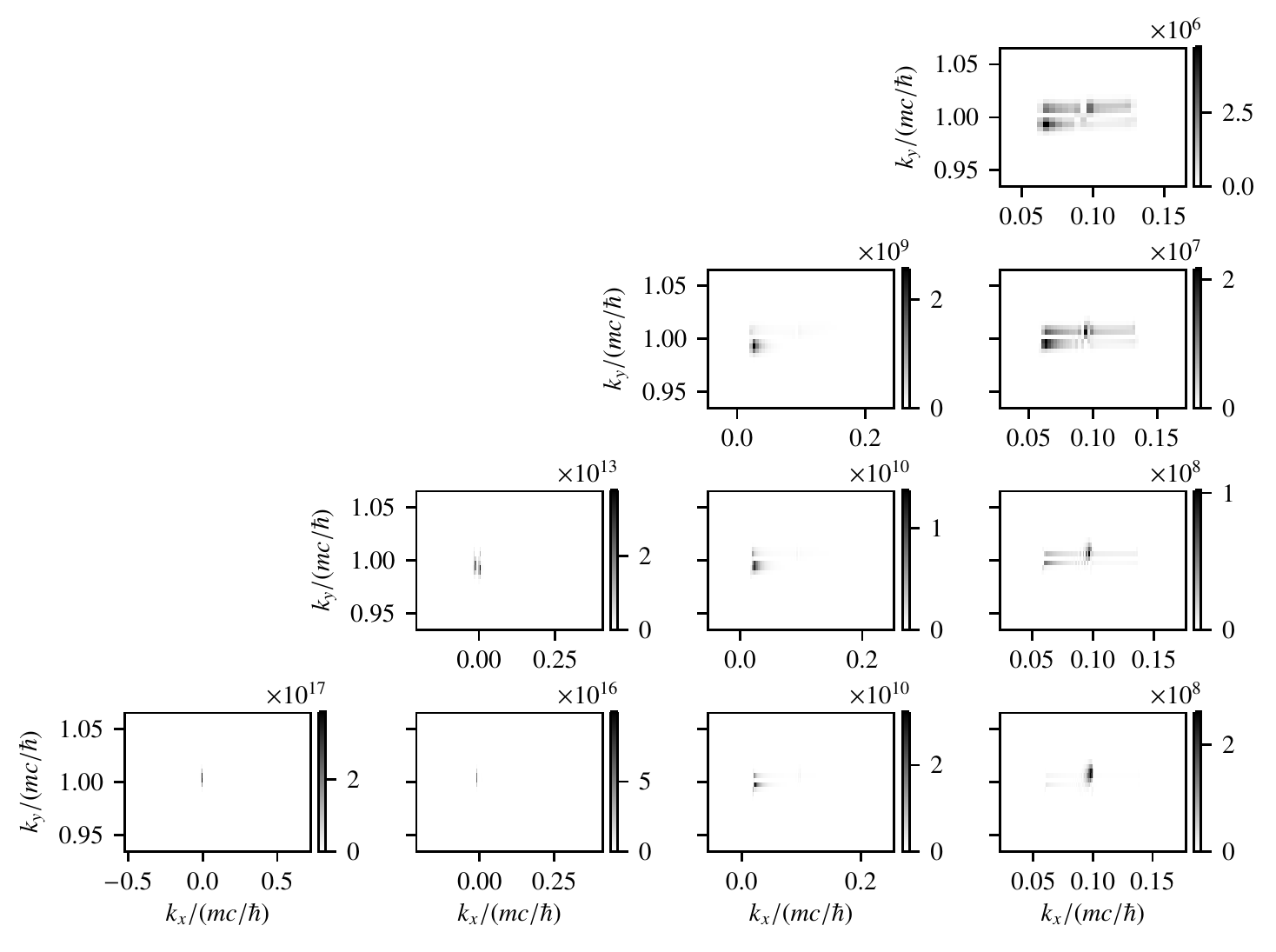}% Here is how to import EPS art
\caption{\label{fig:x_grid_variation_up_zero_longitudinal}Spin-up variation of grid resolution along the $x$-axis without longitudinal polarization component. Shown are the same simulations as in Fig. \ref{fig:x_grid_variation_up}, but with longitudinal vector potential component \eqref{eq:gaussian_beam_exponentials_longitudinal} of the Gaussian beam set to zero. Since the simulation results in this figure and Fig. \ref{fig:x_grid_variation_up} appear similar, we conclude that the longitudinal polarization component of the Gaussian beam has no significant influence for the spin-up quantum dynamics, for the chosen parameters and within the validity bounds of the discussed numerical method.}
\end{figure}
\end{turnpage}
\end{widetext}

\begin{widetext}
\begin{turnpage}
\begin{figure}[H]
\includegraphics[height=0.7 \textwidth]{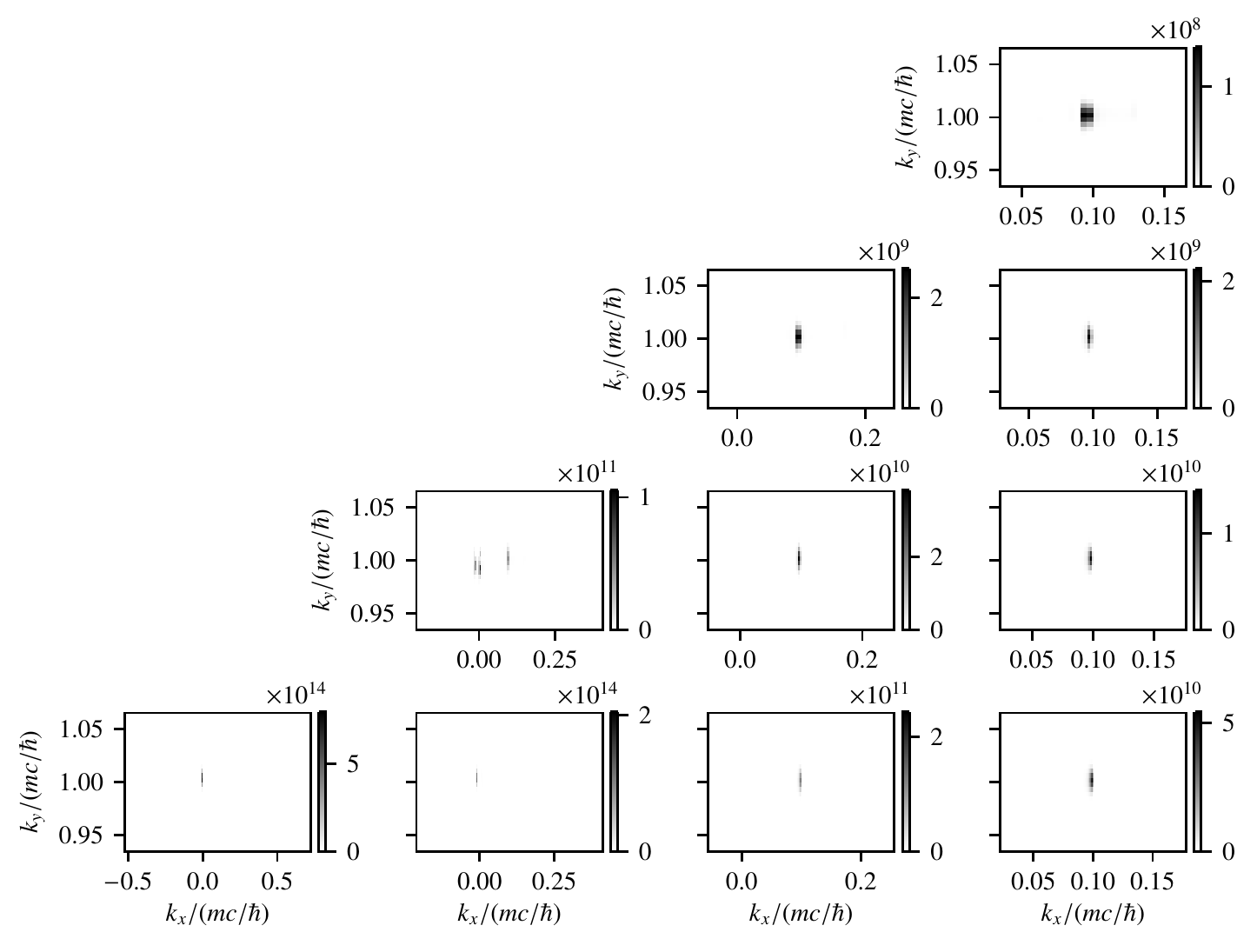}% Here is how to import EPS art
\caption{\label{fig:x_grid_variation_down_zero_longitudinal}Spin-down variation of grid resolution along the $x$-axis without longitudinal polarization component. As in Fig. \ref{fig:x_grid_variation_up_zero_longitudinal}, we display the same simulations as in Fig. \ref{fig:x_grid_variation_down} (spin-down situation), but with longitudinal vector potential component \eqref{eq:gaussian_beam_exponentials_longitudinal} of the Gaussian beam set to zero. We similarly find visual agreement between the present figure and Fig. \ref{fig:x_grid_variation_down}. We therefore conclude a negligibly small influence of the longitudinal polarization component on the spin-down quantum dynamics.}
\end{figure}
\end{turnpage}
\end{widetext}

\begin{table}
\caption{\label{tab:max_values}Maximum values of the diffraction patterns in Figs. \ref{fig:position_y_grid_variation}-\ref{fig:x_grid_variation_down_zero_longitudinal}. The the upper left corner below each separating horizontal lines refers to the figure from which the maximum values are listed. The number alignment corresponds to the panel alignment in the figures, with a $4\times 2$ grid for Figs. \ref{fig:position_y_grid_variation}, \ref{fig:momentum_y_grid_variation} and a lower right triangle in a $4\times 4$ panel grid layout in Figs. \ref{fig:x_grid_variation_up}-\ref{fig:x_grid_variation_down_zero_longitudinal}.}

\begin{tabular}{ c  c  c  c }
\hline
Fig. \ref{fig:position_y_grid_variation}
&  &  &  \\
$4.73076\cdot10^{6}$ & 
$9.29161\cdot10^{7}$ & 
$1.62460\cdot10^{9}$ & 
$2.71331\cdot10^{10}$ \\
$1.41232\cdot10^{8}$ & 
$2.76958\cdot10^{9}$ &
$4.88554\cdot10^{10}$ & 
$8.19968\cdot10^{11}$ \\\hline
Fig. \ref{fig:momentum_y_grid_variation}
&  &  &  \\
$4.73076\cdot10^{6}$ & 
$2.28633\cdot10^{7}$ & 
$1.02838\cdot10^{8}$ & 
$4.32826\cdot10^{8}$ \\
$1.41232\cdot10^{8}$ & 
$6.92659\cdot10^{8}$ &
$3.05408\cdot10^{9}$ & 
$1.28135\cdot10^{10}$ \\
\hline
Fig. \ref{fig:x_grid_variation_up}
&  &  & 
$4.73076\cdot10^{6}$ \\
&  & 
$2.58851\cdot10^{9}$ & 
$2.22438\cdot10^{7}$ \\
& 
$3.75877\cdot10^{13}$ &
$1.37869\cdot10^{10}$ & 
$1.16251\cdot10^{8}$ \\
$3.87179\cdot10^{17}$ &
$9.95994\cdot10^{16}$ &
$3.22529\cdot10^{10}$ & 
$3.69578\cdot10^{8}$ \\
\hline
Fig. \ref{fig:x_grid_variation_down}
&  &  & 
$1.41232\cdot10^{8}$ \\
&  & 
$2.56242\cdot10^{9}$ & 
$2.21080\cdot10^{9}$ \\
& 
$1.08182\cdot10^{10}$ &
$3.85083\cdot10^{10}$ & 
$1.44808\cdot10^{10}$ \\
$8.03467\cdot10^{14}$ &
$2.06701\cdot10^{14}$ &
$2.43485\cdot10^{11}$ & 
$5.41303\cdot10^{10}$ \\
\hline
Fig. \ref{fig:x_grid_variation_up_zero_longitudinal}
&  &  & 
$4.68631\cdot10^{6}$ \\
&  & 
$2.54821\cdot10^{9}$ & 
$2.15155\cdot10^{7}$ \\
& 
$3.73827\cdot10^{13}$ &
$1.36739\cdot10^{10}$ & 
$1.01300\cdot10^{8}$ \\
$3.85711\cdot10^{17}$ &
$9.91970\cdot10^{16}$ &
$3.21221\cdot10^{10}$ & 
$2.60105\cdot10^{8}$ \\
\hline 
Fig. \ref{fig:x_grid_variation_down_zero_longitudinal}
&  &  & 
$1.38042\cdot10^{8}$ \\
&  & 
$2.50477\cdot10^{9}$ & 
$2.18021\cdot10^{9}$ \\
& 
$1.04528\cdot10^{11}$ &
$3.79762\cdot10^{10}$ & 
$1.43718\cdot10^{10}$ \\
$7.91921\cdot10^{14}$ &
$2.03670\cdot10^{14}$ &
$2.41659\cdot10^{11}$ & 
$5.39146\cdot10^{10}$ \\
\hline
\end{tabular}
\end{table}

\bibliography{bibliography}% Produces the bibliography via BibTeX.

\end{document}